\documentclass[twocolumn,a4paper]{article}

\usepackage{url}

\usepackage{times}
\usepackage{helvet}
\usepackage{courier}

\usepackage{subfigure}
\usepackage{graphicx}
\usepackage{tikz,pgf}
\usepackage{pgfplots, xstring} 
\usepackage{pgfplotstable}
\usetikzlibrary{shapes,backgrounds,arrows,automata,fit,calc}

\usepackage{color}
\usepackage{soul}

\soulregister\cite7
\soulregister\ref7
\soulregister\pageref7
\soulregister\eqref7
\soulregister\align7

\definecolor{lightblue}{rgb}{.90,.95,1}
\sethlcolor{yellow}

\usepackage{amsmath}
\usepackage{amssymb}

\usepackage{lipsum}
\usepackage{nameref} 

\usepackage{mdframed}
\usepackage[shortlabels]{enumitem} 
\usepackage{marginnote}

\graphicspath{{images/},{/Users/superk/projects/svn2git/SaoPaolo/CEBI/R/},{/Users/superk/projects/Network/SaoPaolo/output/},{/Users/superk/projects/Network/SaoPaolo/CEBI/R/},{/Users/superk/projects/svn2git/SaoPaolo/CEBI/R/}}

\def\Kin{K_{\rm{in}}} 

\def\a{`{\it{i}}' }
\def\ab{`{\it{i}}'}        
\def\am{\mbox{\it{i}}}        
\def\MAll{M_{CS}}     
\newcommand{\mydot}{.}

\newcommand{\m}[1]{\mbox{\it m}_{#1}}
\newcommand{\ms}{\m{s}}

\newcommand{\Expect}{{\rm I\kern-.3em E}}
\newcommand{\Rs}{\mathfrak{R_s}}
\newcommand{\SRC}{ } 

\makeatletter
\newcommand{\notprop}{\propto\kern-1\@ptsize pt \diagup}
\makeatother

\author{Guy Kelman\textsuperscript{1}, Eran Manes\textsuperscript{2,3}, Marco Lamieri\textsuperscript{4}, David Br\'ee\textsuperscript{5}\\[1ex]
  \fontsize{8}{10.2}\selectfont 1 Hebrew University, Jerusalem, Israel (superk@cs.huji.ac.il)\\[-1mm]
  \fontsize{8}{10.2}\selectfont 2 Ben-Gurion University of the Negev, Be'er Sheva, Israel (msemanes@gmail.com)\\[-1mm]
  \fontsize{8}{10.2}\selectfont 3 Lev Institute of Technology, Jerusalem, Israel\\[-1mm]
  \fontsize{8}{10.2}\selectfont 4 Intesa SanPaolo, Milan, Italy (marco.lamieri@intessanpaolo.com)\\[-1mm]
  \fontsize{8}{10.2}\selectfont 5 University of Manchester, Manchester, UK (davidSbree@gmail.com)
}

\date{September 7, 2017}

\begin{document}
\renewcommand*{\raggedleftmarginnote}{\flushleft}


%
%

\title{Missing Data as Part of the Social Behavior in Real-World Financial Complex Systems}

%
%
%

\maketitle

\begin{abstract}
    Many real-world networks are known to exhibit facts that counter
    our knowledge prescribed by the theories on network
    creation and communication patterns.
    A common prerequisite in network analysis is that information
    on nodes and links will be complete because network topologies
    are extremely sensitive to missing information of this kind.
    Therefore, many real-world networks that fail to meet this
    criterion under random sampling may be discarded.

    In this paper we offer a  
    framework for interpreting the missing
    observations in network data under the hypothesis that these
    observations are not missing at random. We demonstrate the
    methodology with a
    case study of a financial trade network, where
    the awareness of agents to the data collection procedure
    by a self-interested observer may result in strategic
    revealing or withholding of information.
    The non-random missingness has been overlooked despite the
    possibility of this being
    an important feature of the processes by which the network is generated. 
%
%
%
    The analysis demonstrates that strategic information withholding
    may be a valid general phenomenon in complex systems. 
    The evidence  is sufficient to support the existence of an
    influential observer and to offer a compelling dynamic mechanism
    for the creation of the network.

\end{abstract}


\section*{Introduction}
\label{sec:intro}


\noindent
  When individuals are aware of being monitored, their behavior is
  likely to improve. This intuitive insight was first studied
  through a sequence of experiments in the field of organizational
  behavior,  conducted at the Hawthorne works in the 1920s
  \cite{Franke:1978vn}. The trials showed that the workers' mere
  awareness of being monitored resulted in increased productivity,
  regardless of the direction of change in the working conditions.  This effect
  is known as the `Hawthorne Effect', or the `Observer Effect'.

  In this paper we argue that the presence
  of a self-interested data collector in financial complex networks
  may instigate a
  change in the behavior of participants such that the network visible
  to the collector differs from the actual one. This effect
  has far-reaching implications on our
  understanding of the  processes generating such networks:
  the way by which the players connect or dissociate, the
  kind of information that is hidden, and the overall effect of
  this on the stability
  and functioning of these systems.


  Imagine for a moment that a monitoring entity such as a federal
  or regulatory agency, a bank or even a concerned parent, is
  observing a network of social or business ties
  (a crime network, the financial transactions of a stock broker,
  or the social activity of a child in his favorite social network).
  Assume further that subjects are
  aware of this monitoring procedure. It is only natural to expect
  that some information will be withheld, strategically, from the
  monitoring entity. The strategic disclosure of information may
  be a consequence of the activity in the neighborhood where the
  agent resides, and the sensitivity of his payoff to the observer's
  presence.
  Unfortunately, detection of the influential presence of an observer
  is extremely challenging because the true underlying network is
  not given to us, and thus it is particularly hard to discriminate
  the true dynamics from the visible one.
  Notwithstanding, we believe that this strategic withholding of
  information from the observer leaves traces on the visible topology
  as well as on other, intrinsic, measurable attributes of the
  participants.


  Although it is possible that many complex systems in the
  real-world are influenced by the data collection procedure,
  only a small subset fulfil the set of prerequisites which may
  offer the opportunity to test the hypothesis that the visible
  system was formed under the presence of an influential observer.
  The prerequisites are:
  \textbf{(i)}
  The network appears to be dominated by sampling error, with abnormally large number of disconnected components, large diameter, or a small number of neighbors on average.
  \textbf{(ii)} There exists an underlying theory that specifies the
  motivations of different agents to hide or disclose information,
  and the interest of the observer to monitor the ties formed by
  them.
  \textbf{(iii)} The consequent sensitivity of different agents
  to the observer's presence is quantifiable and could be retrieved from
  the data.
  \textbf{(iv)} The visible network permits the measuring of the
  intrinsic attributes of agents, the communication patterns between
  them, and the incentives to disclose or withhold information.
  To this list we could add the trivial requirements that is met by most network models:
  \textbf{(v)} It is possible to obtain topological aggregates of
  the visible network, and to compare them with the topological
  measures of a network under full information.

  We propose a procedure to measure a system under monitoring. The procedure begins by analyzing the topological attributes of the visible network vis a vis the expected structure under full information. If major discrepancies appear to be present, we move on as follows: we identify the observer and the agents, classify agents into classes (species) that exhibit different sensitivities to the presence of the observer, and prescribe an information disclosure strategy to each class. The proposed classification is derived from a known underlying theoretical foundation (e.g. profit maximization, risk aversion, thinking at the margin) that may support such differential information disclosure strategy. We then check whether the visible network indeed exhibits such species-dependent information exposure - as predicted by the theory - and compare the revealed effect with the one expected under full information. We complement the procedure by adding mediation analysis to rule out possible alternative dynamic mechanisms of the system that are observationally equivalent.


  To make our point we searched for a complex system that has
  interacting agents, and where a collector of the data possesses
  potential impact on the payoffs of the agents. We further required that it
  would be possible to estimate the sensitivities of agents to the
  observer's presence.
  Naturally, financial networks admit to these criteria. These
  networks usually consist of profit-maximizing agents (firms) that
  are grouped into well defined classes according to attributes
  such as industry, credit-rating, firm size, aversion towards risk, liquidity, and
  so on. Moreover, data on these networks are normally recovered by
  self-interested third party agencies that interact with subsets of the
  agents and possess a direct and measurable influence on the agent's
  payoffs. These self-interested parties may be banks or regulatory
  agencies.

  We were able to locate a large-scale business
  network where firms are buyers and sellers and the transactions
  between them are the ties. These data are collected by a bank for
  operational needs of maintaining credit lines with its borrowers.
  Thus it plays the role of the observer and has a clear incentive
  to monitor the financial activity of its customers. This network
  complies with the prerequisites prescribed above; it is in the
  real world, the presence of the bank may motivate the agents to strategically hide
  or disclose information, and all the intrinsic attributes of the
  agents are quantifiable, allowing us to control for a rich set
  of covariates, e.g. from firms' financial statements. And last, the
  topology of the network may provide additional insight that may
  help to uncover
  the processes generating such networks. The generating processes can unravel if we could use the network to discover production chains, or rather production trees by means of flow dynamics, or locate central nodes by means of agent-based traversal.

  Many studies of financial networks \cite{Tamura:2012ys},
  \cite{Sieczka:2011zr}, \cite{Boissay:2006ix} assume that the
  network data are complete in the sense that the visible nodes and
  links are all there is. Such a working assumption is shared across
  disciplines that use networked data, and the sampling
  procedures are usually random in order to correct for known biases
  \cite{Rosenbaum:1983dq}, \cite{Huisman:2003fk}, \cite{Huisman:2009fk}.
  Yet, what if links are missing intentionally, that is, not at
  random? And what if the process that is causing linkages to be
  missing is itself a process of the network?
  With these questions in mind it may be plausible to revisit the
  family of works on trade credit \cite{Omiccioli:zr}
  \cite{Petersen:1997kx} and others on stock exchange trading and
  modeling 
  noting that another kind of process that causes missing data
  may exist; and that this process is part of the network's generating
  mechanisms.
  We believe that our approach
  offers a change in perspective in the analysis methodology
  that may constitute a proper
  statistical examination for this possible non-random cause of
  missingness,  non-response \cite{Stork:1992cr},
  \cite{Huisman:2008uq} and non-observability \cite{Rosenbaum:1983dq}.

  Measuring real-world complex networks has become a growing trend in the
  past decade \cite{Huberman:1999fk}, \cite{PhysRevE.64.046132},
  \cite{citeulike:458130} mainly for two reasons: \textbf{(1)} the development
  of methods for deducing
  the processes underlying the generation of a network from its structure \cite{Redner:1998sj} \cite{Dorogovtsev:2000eq}
  \cite{PhysRevE.64.025102} and \textbf{(2)} the growing availability
  of detailed digital data on the interactions between participants.
  However, the overall impact of many such  underlying processes
  deduced from  network structures has been  called into question
  because of a suspicion that, in the process of creating a network
  model, links were removed in the sampling process.
  The accidental removal of even a single link may, in some cases,
  deform the realized network's topology such that the
  researcher is led to interpret the structure inaccurately
  and create a false understanding of the real processes.
  Furthermore, in networks it is sometimes useful to predict the
  missing links in order to obtain a workable structure
  \cite{Liben-Nowell:2007fk}, \cite{Adamic:2003uq}. Now, link
  prediction requires knowing the class of network under investigation.
  False inference on the generating process may lead to a
  wrong guess of the network classification and may pull the
  prediction to form a network model that is not true to reality.
  It is therefore not surprising that a myriad of such `damaged'
  networks are, and were, discarded.

  In a later study, \cite{Guimera:2009aa}, the authors
  address the problem of the reliability of links. They use a stochastic block model framework to create a reliability measure of individual links, given the observation and a family of acceptable realizations thereof. The link reliabilities are used to identify missing and spurious interactions in the observed network. The authors then test the performance of their approach by randomly adding and removing links from five high-quality, error-free networks which results in high accuracy of feature recovery.
  While their approach is designed to predict the potential appearance of spurious links,
  our methodology aims to provide non-invasive tools that would assist in determining whether nodes or links are missing at random, or that a strategic, systematic missingness is more likely the case.
  For this reason, we stress that this study does not aspire to offer a method for link prediction or draw inferences on the topology of the true network.


  Our paper is related to several other distinct branches of existing
  literature. One is
  on-line self-disclosure and impression management, and information
  manipulation in social networks. This literature demonstrates the
  prevalence of information manipulation regarding participants’
  physical attributes in on-line dating sites \cite{Toma:2008uq}, and the
  role of the number of friends and their appearance and behavior
  on evaluations of individuals in social networks
  \cite{Utz:2010fk}, \cite{Tong:2008kx}, \cite{Walther:2008vn}. Further, it relates to discussions on the strategic, signalling role of displaying
  one's social connections on social network profiles
  \cite{Donath:2007ys}, \cite{Donath:2004zr}. The main empirical finding in these papers is that
  popularity and attractiveness of individuals in social networks
  are strongly affected by their friends' appearance
  . While our
  study finds similar effect in a real world financial network it has a different focus and implications.
  Popularity, which determines social status and payoff, is replaced by credit worthiness,
  that determines financial costs and so economic profits.
  Rather than focusing on appearance of one's friends' and the effect
  it has on the social status, our work is centered around developing
  a method to uncover strategic information withholding given the varying
  sensitivities of different agents' payoffs to the appearance of
  their social or business ties.
  Our main claim is that manipulated information has an
  effect on the part of the network that is visible, rather than
  on the status of individuals, and this is what we seek to explore.

  Further, the paper is related to the literature on hidden populations.
  This literature highlights the usefulness
  of networks in resolving the sampling problem of subgroups that are
  deliberately missing from the records \cite{Rubin:1976kx},
  \cite{Frank:1994fk}, \cite{Huisman:2008uq}.
  Several studies are concerned with problems of missing data in
  networks in longitudinal studies. There, evolution
  of the network structure is part of the cause for missingness
   \cite{Huisman:2003fk}.
  Sampling and consequent analyses of hidden populations, such as
  cocaine users, sexual disease transmission, criminal networks or
  other networks where observations are missing, may be carried out
  by link-tracing \cite{Pattison:2013uq}, \cite{Simsek:2008kx} or
  link-prediction \cite{Clauset:2008kx}, \cite{Ohnishi:2009uq}.
  Alas, in many cases link tracing is not possible since a second round of queries to the actors may not be attainable, and link prediction should be avoided for reasons mentioned above.
  Some argue that more data in greater detail are needed to accurately
  derive the nature of the social effect \cite{Manski:1993fk}.
  However, if data are missing not at random gathering more data will not improve the network description.

  In a recent paper directly related to ours \cite{Smith:2017fk}, the authors make a first systematic  attempt to account for non-random missingness of links and its effect on estimates of  key network statistics, by removing nodes through a weighting process that factors in centrality of nodes and chance. Using select networks and  network measures, and controlling for the portion of missing links, the authors  then pursue a set of Monte Carlo simulations of taking out nodes in these networks. Finally, they compare the simulated to the original ones in order to gauge the  level of bias that missing links impose. Their main finding is that bias is worse  when central nodes are missing. While their approach is intended to gauge the level of bias, given the percentage of links  missing and species (e.g. central vs. non-central), our contribution is to determine how likely is the given visible network formed through a process of strategic withholding of links from an influential observer, without any prior knowledge on the node species and percentage of  missing links.


  The rest of the paper is organized as follows; the \textbf{\nameref{sec:methods}} section 
  gives basic definitions and formal concepts, and describes the procedure.
  In section \textbf{\nameref{sec:example}} we apply the methodology
  on a large-scale real-world inter-firm trade network. 
  We conclude and 
  discuss several implications
  and limitations of our method in the last two sections. In the appendices we give an overview
  of financial trade networks, and lay out descriptive statistics that
  accompany the case study.
  \section*{Materials and Methods}
  \label{sec:methods}
  In this section we lay out the formalism of two mutually dependent
  network models: the \emph{monitored} network $(V,E)$ of $N$ nodes
  $V=\{i_1,\dots, i_N\}$ and $L$ links $E = \{e_1, \dots, e_L\}$
  and the \emph{true} network $(V',E')$ that has $N'$ nodes and $L'$ links.
  The nodes perform interactions, and so they will be referred as \emph{agents}.
  Further, a single, unique, observer exists in the system, who can
  see the monitored $(V,E)$ but not the true network $(V',E')$.
  This observer interacts with  agents and may reward them based
  on the agents to which they are linked.
  We assume that the visible network is a subset of the true one. Formally we write
  $(V,E) \subseteq (V',E')$, and consequently $N \le N'$ and $L \le L'$.

  We mark the symbols designating the true network with a prime ($^\prime$).

  \subsection*{Basic Definitions}
  \label{sec:basicdefs}
   We consider the network of $N'$ agents, of which only $N$  are
   visible to the observer. A directional weighted link exists between  pairs of agents if they interact.
   An ordered pair $(c,s) \in V \times V$ designates a directional
   link between an interacting pair of two nodes $c, s \in V$, who
   play different roles in this interaction. Each directional link
   $c\rightarrow s$ has an associated weight $w_{cs}$.  Similarly
   for the pair  $(c',s') \in V' \times V'$. To motivate this we
   could imagine a network of (c)ustomers who pay amounts of $w_{cs}$ upon purchasing goods from (s)ellers.

   Let $\Theta$ be the set of all possible agent species
   with a typical element $\theta \in \Theta$.
   The agent's \emph{species} is relevant in our set-up, as the observer determines an agent's pay-off based on their species.

   Following standard Bayesian analysis, there exists a prior
   probability distribution $p$ over the set $\Theta$ which is common knowledge,
   such that $\Pr(\theta \in \Theta)$ is known.
   The categorized species of an agent could be its credit rating, as discussed in the example section \textbf{\nameref{sec:example}}.
   We mark the species of agent $s$ by $\theta_s$.
   It is noteworthy that $\theta_s$ may not be accessible to the observer, and that this fact will come in handy later on.
   In a similar way we can
   define the bandwidth function $W_s$ that may operate on the subset of agents $X$ neighboring with $s$ to
   extract the weights on the links pointing to agent
   $s$: $W_s[X] = \{w_{xs}\ |\ (x,s) \in X\times V\}$.
   We will use $W$ instead of $W_s$ where it is obvious.

   The pair $(c',s')$ in the true network, may not
   be visible to the observer. This can  happen only if the pair
   is in the complement of the visible network $(V^\dag, E^\dag)$
   \begin{align}
     (V^\dag,E^\dag) &= (V',E') \setminus (V,E)
     \label{eq:VE}
   \end{align}
   Pairs that are symbolized $(c,s)$ are strictly members of the
   visible network and ones written as $(c^\dag, s^\dag)$ are strictly invisible to the observer.

\begin{figure}[htb]
  \centering
    \begin{tikzpicture}[->,>=stealth',shorten >=1pt,auto,node distance=2.8cm,semithick]
        \tikzstyle{firm}=[state,fill=blue!80,draw=none,text=white,scale=0.5,node distance=2.8cm]
        \tikzstyle{buyer}=[firm,node distance=1.8cm]
        \tikzstyle{seller}=[firm,node distance=3.5cm]
        \tikzstyle{outside buyer}=[buyer,fill=none,draw=black,text=black]
        \tikzstyle{outside seller}=[seller,fill=none,draw=black,text=black]
        \tikzstyle{hidden}=[firm,fill=white,draw=none,text=black]

        \node[buyer] (c1)                    {c1};
        \node[buyer] (c2) [above of=c1]      {c2};
        \node[buyer] (c3) [above of=c2]      {c3};
        \node[seller] (s1) [right of=c2]      {s1};

        \node[seller] (s2) [below right of=s1]      {s2};
        \node[buyer] (c4)  [right of=s2,node distance=3.5cm]  {c4};
        \node[buyer] (c5) [below of=c4]      {c5};
        \node[buyer] (c6) [above of=c4]      {c6};

        \node[outside buyer] (c7) [below left of=c5]      {c7};

        \node[outside buyer] (c8) [below of=s2,node distance=3.5cm]  {c8};
        \node[outside buyer] (c9)  [left of=c8,node distance=1.5cm] {c9};

        \node[outside seller] (s3) [left of=c1,node distance=2cm] {s3};
        \node[outside buyer] (c12)  [left of=s3, node distance=2.7cm] {c12};
        \node[outside seller] (c11)  [below left of=s3, node distance=2.7cm] {c11};
        \node (c13) [outside buyer] [above left of=s3, node distance=2.7cm]{c13};
        \node (c10) [outside buyer] [below of=s3,node distance=2.7cm]      {c10};

  \begin{scope}[every node/.style={scale=.6}]
    \path   (c1) edge [bend left=20]   node{$w_1$} (s1);
    \path   (c2) edge [bend left=20]   node{$w_2$} (s1);
    \path   (c3) edge [bend left=10]  node{$w_3$} (s1);
    \path   (c4) edge [bend right=10]   node{$w_4$} (s2);
    \path   (c5) edge [bend right=20]   node{$w_5$} (s2);
    \path   (c6) edge    node{$w_6$} (s2);
    \path   (c7) edge [dashed]  node{} (s2);
    \path   (c8) edge [bend right=20,dashed]   node{} (s2);
    \path   (c9) edge [bend right=10,dashed]   node{} (s2);
  \end{scope}

    \path   (c13) edge [dashed] node{} (s3);
    \path   (c12) edge [dashed] node{} (s3);
    \path   (c11) edge [dashed] node{} (s3);
    \path   (c10) edge [dashed] node{} (s3);
    \path   (c10) edge [bend right=20,dashed] node{} (s1);

  \tikzset{blue dotted/.style={draw=blue!50!white, line width=1pt,
                               dash pattern=on 1pt off 4pt on 6pt off 4pt,
                                inner sep=1mm, inner ysep=1mm,
				rounded corners=2mm, rectangle}};

  \tikzset{no border/.style={draw=none,
                                inner sep=0mm, inner ysep=1mm, rectangle}};

  \node (first dotted box) [blue dotted,
                            fit = (s1) (s2) (c1) (c2) (c3) (c4) (c5) (c6) ] {};
  \node (O) [outside buyer,double,anchor=north east,yshift=-5mm,xshift=-5mm] at (first dotted box.north east) {O};

  \begin{scope}[every node/.style={scale=.7}]
    \path (s1) edge [-latex',double, =>,shorten >=3pt,semithick, bend left=20] node {$\m{s1}$} (O);
  \path (s2) edge [-latex',double, =>,shorten >=3pt,semithick,bend left=10] node {$\m{s2}$} (O);
  \end{scope}

  \node (second box) [no border,
                            fit = (c7) (c8) (c9) ] {};

  \node (third box) [no border,
                            fit = (s3) (c10) (c11) (c12) ] {};

    \end{tikzpicture}
    \caption{\small A sketch of the possible situations in a
    monitored network. Following (\ref{eq:VE}), the boxed area is $(V,E)$, where
    information on the agents and the links is visible. Dashed links
    and hollow nodes outside of the boxed area are in $(V^\dag,E^\dag)$.
    The blue agents are either in contact with the observer $O$
    or exposed to it. The double arrows carry messages between
    the agents and the observer. In
    the text are some walk-through examples.}
  \label{fi:schematic}
\end{figure}
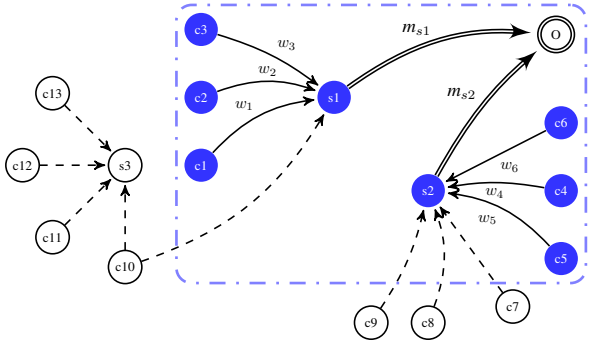

   We define the set of actions available to the observer as $A=\{a_1,
   a_2, \dots\}$ with a typical action $a_o\in A$
   . Similarly, we define by $M$ the set of actions available
   to the agent, and call this set the \emph{messages}.  A
   message $\ms \in M$ sent out by agent $s$ gives away part of the information
   about agents that are directly and \emph{visibly} linked to $s$, $\mathfrak{V}_s$
   \begin{align*}
     \mathfrak{V}_s &= \left\{ c \ |\ (c,s) \in (V,E) \right\}
   \end{align*}
   In contrast, the true neighborhood of $s$, $\mathfrak{V}_s'$, may encapsulate $\mathfrak{V}_s$ and can be defined as
   \begin{align*}
     \mathfrak{V}_s' &= \left\{ c \ |\ (c,s) \in (V',E') \right\}
   \end{align*}
   Examples of these definitions can be seen in figure \ref{fi:schematic}: $\mathfrak{V}_{s1} = \{c1,c2,c3\}$, $\mathfrak{V}_{s1}' = \{c1,c2,c3,c10\}$, $\mathfrak{V}_{s2} = \{c4,c5,c6\}$, and $\mathfrak{V}_{s2}' = \{c4,c5,c6,c7,c8,c9\}$. $s3$ and its neighborhood $\mathfrak{V}_{s3}$ are all in $(V^\dag,E^\dag)$. The messages passed to observer $O$, $\m{s1}$ and $\m{s2}$, can be any set of parameters on $\mathfrak{V}_{s1}$ and $\mathfrak{V}_{s2}$.

   We now focus on the right member of the ordered pair, $s$, who we define as
   the agent that may interact with the observer.  This member of
   the pair may choose to play strategies in which partial or all information may be concealed from the
   observer.
%
  For this reason, the species of agent $s$ has two components: a historical
  component $\Rs$ based on past interactions with the observer, and
  the social interaction component $W[\ms']$


  The reason for this separation is that $\Rs$ is accessible to the observer at all times,
  but the neighborhood-related component $W[\ms']$ that potentially affects
  the payoffs, may not be so. To illustrate the use of $\Rs$ and
  $W[\ms']$ we could consider the credit rating $\Rs$ of a firm $s$
  which reflects the historical moral of payments of the borrower.
  However the overall risk of default might also depend on the current
  business activities with the financial neighborhood of first
  degree. Thus $W[\ms']$ may be the mean financial costs among all
  the trade-partners. Hereafter we define $W_s' := W[\ms']$ for
  brevity and call it the \emph{neighborhood indicator} of agent $s$.

   Asymmetry of information thus exists owing to the fact that $\theta_s$
   is selectively disclosed and should be estimated by the observer
   from a probability distribution. So, in order to take proper
   action the observer estimates the species
   of $s$ based on this prior probability $\Pr(\theta_s \in \Theta)$ and
   additional information delivered by the message $\ms$.

   A possible message may contain the identities
   of all the neighbors of $s$:
   \begin{align}
     \ms' &= \left\{ c\ |\ c\in\mathfrak{V}_s' \right\}
     \label{eq:minimalMS}
   \end{align}
   however, the agent
   may play a strategy of disclosing partial information about
   his neighbors. i.e. sending the message $\ms \subseteq \ms'$. 
   A message that is a subset of $\ms'$ we name the \emph{information set} of node $s$.

   In several set-ups we may find that more information is carried than only the set of neighboring nodes. We will encounter such a set-up in the example section. Thus \eqref{eq:minimalMS} may be redefined as
  \begin{align}
    \ms &= \left\{ (c,\hat{\theta}_c)\ |\ c\in \mathfrak{V}_s \right\}
     \label{eq:comprehensiveMS}
  \end{align}
  for a subset of $\ms'$, where $\hat{\theta_c}$ may be any set of arguments intrinsic to $c$. The \textit{information exposure} of each node could be a summary indicative of the difference between $\sum_c \hat{\theta_c}$ and its counterpart $\sum_c \hat{\theta_c'}$ under full information.

   The payoff that every player would like to maximize in his own
   benefit is a scalar marked by $u(\cdot)$. Payoffs are earned after each single binary interaction of the observer and an agent. There are three determinants to the payoffs:
   the agent species, its message and
   the observer's action. This can be written as
   \begin{align*}
     u_s,u_o : \Theta \times W[M] \times A \longrightarrow \mathbb{R}
   \end{align*}
   where the payoff of an agent is $u_s$, and the observer's payoff
   from acting upon is $u_o(s)$. Here, $W[M] = \{W_s | \ms \in M\}$
   is the set of all neighborhood indices estimated from the set
   of messages $M$.

  Let us now recall a previously drawn
  distinction between  monitored and  unmonitored networks.
  Designation of the social ties corresponds, in general, with two
  dimensions: the species $\{\theta_c\ |\ c \in \ms'\}$ and the species $\theta_s$ of the supplier $s$, and the weights on the links pointing to $s$,  $W_s'$. 
  With perfect information, the observer may take action
  $a'$ corresponding exactly to these parameters, to receive the
  optimal payoff in a transaction with agent $s$, $u_o'(s) = u(\theta_s, W[\ms'], a')$. If
  information is reduced, the observer must complement the missing
  details with what he has. The optimal payoff then becomes
  $u_o(s) = u(\hat{\theta}_s, W[\ms], a)$, a function
  of the reduced message $\ms$, the expected
  species of the agent $\hat{\theta}_s$, and the now optimal action
  $a$. Hereafter in certain contexts we may omit the factors that
  $u(\cdot)$ implicitly relies on.

  For two individuals $s_1$ and $s_2$ of different species $\theta_{s_1} \ne \theta_{s_2}$,
  we say that $s_1$ is more \emph{sensitive} than $s_2$
  to the impact of its relationship with its neighborhood
  whenever its expected payoff 
  is more responsive to the neighborhood indicator. Namely, 
%
  \begin{align}
    \frac{ d \Expect\left[ u(W_{s_1}')\ \big|\ \theta_{s_1} \right]}{d W_{s_1}'}
     >
     \frac{ d \Expect\left[ u(W_{s_2}')\ \big|\ \theta_{s_2} \right]}{d W_{s_2}'}
     \label{eq:neighborhoodIndex}
  \end{align}
  from this follows the information disclosure strategy of the
  species towards the
  observer in the sense that $s_1$, being of species $\theta_{s_1}$ is more
  selective in exposing information.
  This condition motivates our conception of measuring the network
  under the hypothesis of an influential observer.

  The next section uses these diagnostics in a case study of a
  real-world business network. There,
  we classify the agents to species in the form of credit-rating classes,
  we identify the self-interested observer in this network which is the bank,
  we show the sensitivity that firms of different rating classes exhibit to the
  impact of the bank's rating mechanism, and that given this differential sensitivity
  we define the information exposure that
  gives rise to two main strategies of disclosure: as expected,
  agents that are more sensitive to their neighborhood's trading pattern will be
  more selective in giving away information to the bank.
  We explain these by means of
  standard economic theory, and conclude with mediation analysis
  to rule out alternative hypotheses to the existence of a monitoring
  entity.

  For general background on financial networks, and descriptive statistics
  on our network, collected by an Italian bank 
  the reader is referred to appendices \ref{app:financialNetowrks} and \ref{app:stats}.


   \section*{A case study in the real-world}
   \label{sec:example}

   In this section we deploy analysis on a test case from the financial real world. We follow the procedure prescribed in the introduction, namely:
   \begin{itemize}
     \item Present evidence on network attributes with contrast to what would be expected under full information.
     \item Describe a theory from which a species-dependent information exposure is likely to be present.
     \item Present the revealed pattern in our test case network compared to the one expected under random sampling design.
     \item Perform mediation analysis where we rule out the possibility that this observation is given by other intervening variables.
   \end{itemize}

   Let us now consider the financial system that was mentioned
   previously, where the agents are firms that buy or sell from
   each other using invoices (later to be termed \emph{Trade Credit}), and the observer is a bank.  The
   theoretical foundation
   is the profit maximization principle \cite{Varian:1992uq}, the
   species are risk classes of agents, the payoffs
   are the profit functions of the firms, the messages that firms
   send are 
   the result of a strategic choice on the amount of sales invoices to expose
   and which of their sales invoices to bring to the bank as collateral for a loan,
   and the strategies that the bank plays are the
   interest rates charged on their loans. 
   Both the agent and the observer
   play their strategies conditional on the species  of the
   agent.

   We will soon find that many observations on this network place in
   our hand the possible rejection of the hypothesis that the network
   is unmonitored. A brief, non-exhaustive, list of evidence is the elimination of
   most nodes and links with minimal cleaning procedure, the complete
   breakage of the network to paths shorter than the production
   chain \cite{Kelman:2015fk}, and the inability to detect propagation
   of distress in the 2008 financial crisis \cite{Golo:2015uq}.

   For further details on financial networks in general and trade credit networks we refer the reader to
 appendix \ref{app:creditRatingCosts}.

  \subsection*{Network characteristics}
  \label{sec:networkCharacter}

  In order to convince the reader that the network has a problem we intend to contrast several
  stylized facts: \textbf{(a)} the network is directed and is an agglomerate of trees (such that money is expected to travel long distances), \textbf{(b)} the network is negatively assorted (so that agents of similar connectivity don't engage), \textbf{(c)} the structure of the system is similar to a social network, and:
  \textbf{(d)} money travels short distances (long paths are not utilized), \textbf{(e)} buyers are also sellers (the network is not bipartite and loops \ul{should} occur), and \textbf{(f)} the agents have a strong tendency to affiliate with like agents (contrasting the negative degree assortativity feature).

  Taken together, bullets (a) through (f) surrender the possibility that the network indicates that an influential observer exists.
  This concludes the first prerequisite and warrants further testing according to our proposed methodology.

   \renewcommand{\arraystretch}{1.3}
   \begin{table*}[h]\par
  \scriptsize\centering
  \begin{tabular}{c|c|c|c|c|c} 
    \textbf{network} & \textbf{email} & \textbf{LJ} & \textbf{WikiTlk} & \textbf{Amazon} & \textbf{Our} \\ 
\hline
Nodes              &  265214        &  4847571         &  2394385        &  403394         &   345403         \\  
Edges              &  420045        &  68993773        &  5021410        &  3387388        &   2874830        \\  
Nodes Largest WCC    &  224831 (0.85) &  4843953 (0.99)  &  2388953 (0.99) &  403364 (1.00)  &   341023 (0.98)   \\  
Edges Largest WCC    &  395270 (0.94)  &  68983820 (1.00) &  5018445 (0.99) &  3387224 (1.00) & 1091873 (0.38) \\
Nodes Largest SCC    &  34203 (0.13)   &  3828682 (0.79)  &  111881 (0.05)  &  395234 (0.98)  &   101186 (0.29)   \\
Edges Largest SCC    &  151930 (0.36)  &  65825429 (0.95) &  1477893 (0.29) &  3301092 (0.97) &  193198 (0.56) \\
Clustering Coeff    &  0.0671        &  0.2742          &  0.0526         &  0.4177         &   \ul{0.003}     \\  
Triangles          &  267313        &  285730264       &  9203519        &  3986507        &   195318         \\
Frac. Closed. Triangs  &  0.001373      &  0.04266         &  0.001112       &  0.06206        &   \ul{2.04e-05}  \\
Diameter           &  14            &  16              &  9              &  21             &   20             \\  
90\% Effective Diam.  &  2.5           &  6.5             &  4              &  7.6            &   12             \\  

\hline
      \end{tabular}
      \caption{\small Comparison of topological measures in directed networks that may possess a social component. The parentheses give proportions from the full data. Our network is on the rightmost column, other data sets are from \cite{Leskovec:2014aa}. The clustering coefficient and the fraction of closed triangles (underlined) indicate a directional tree-like structure, which is expected from a production chain. However, together with the large diameter it contrasts the reciprocity of roles, namely that half of the nodes are buyers \ul{and} sellers. WCC=Weakly connected component, SCC=Strongly connected component.}

  \label{tab:SNAPandMore}
\end{table*}

  We begin our exploration by looking at some bare topological measures of our system and comparable ones. Table \ref{tab:SNAPandMore} lists networks that are known to be directed and possess a social component. Following a cleaning procedure that utilizes financial considerations, we present our network on the rightmost column. We elaborate on the cleaning procedure in appendix \ref{app:stats}.

    Browsing the common measures of the directed network, we find that connected component sizes are small.
    The size of the largest component is 101,186 nodes (out of a total of 345,403 nodes) with a diameter of 20. Of the remaining 244,217 nodes, 239,780 are situated in small clusters of 1 link each.  In many of the sub-graphs the central nodes deal small contract sizes, e.g. phone companies or couriers, each of which is financially irrelevant to the system due to the extremely small total transaction volumes between them and each of their peers.  Filtering out these irrelevant firms causes the network to break completely.

    Next we address the question of whether the network is similar to a social network, from interaction of
    seller-buyer pairs and their respective credit ratings.  The distribution
    of the in-degree of sellers often reveals information about the
    way a network is formed. Here  the in-degree distribution is a power law, indicating
    an association mechanism similar to popularity. Although
    the seller is required to recruit the buyers, in the relevant literature
    on diffusion we witness an inevitable
    coupling of mass media (external factors) and word of mouth
    (local factors). A good historical review of market models can
    be found in \cite{Goldenberg:2000uq}.
    In order to render a power law of the degree distribution, the
    seller's own reputation should act to reverse the damping effect of
    diffusion due to e.g. advertisement. This confirms the point
    we made that the network is social.

\begin{figure}[htb]
   \centering
   \subfigure[mass]{
   \includegraphics[scale=0.31,bb=10 30 512 450]{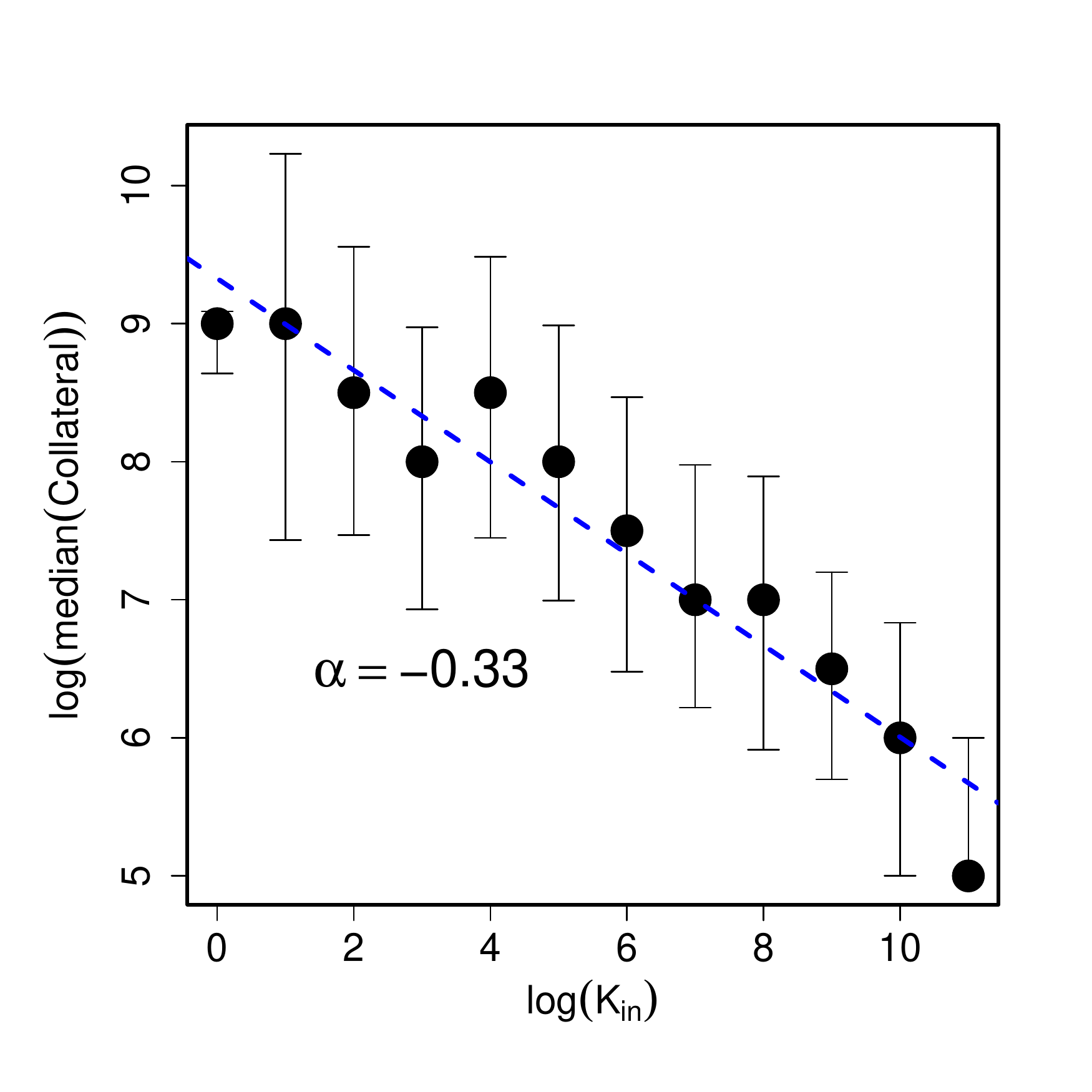}
   \label{subfig:kvsPji}
   }\hfil
   \subfigure[number]{
   \includegraphics[page=3,scale=0.3,bb=10 10 512 490]{run\mydot UO4m3/knn_CDF\mydot in\mydot 0_00\mydot UO4m3.pdf}
   \label{subfig:knn}
   }
   \caption{\small
   Median contract size \ref{subfig:kvsPji} and average buyer
   (neighbor) degree \ref{subfig:knn} plotted
   against seller's degree $\Kin$,
   from the sellers in the trade-credit full network (
   n=273,726).
   The network is dissortative both in connectivity and mass.
The slope in \ref{subfig:knn} can be approximated by $\Kin^{-1.24}$.
   \SRC}
   \label{fig:kvsPji}
\end{figure}

    As for peer-interaction in 
    financial context, the credit rating of sellers and their buyers are measurables external to the network's topological ones. Using these, we intend to show more evidence supporting our claim of non-random missingness:
    Generally, large firms deal with large contracts, small firms deal small.
    Although there are exceptions, e.g. the phone and courier companies,
    figure \ref{fig:kvsPji} indicates that having many
    buyers means both that the median contract size is small and
    that the neighbor's degree is likely to be small. 
    The conclusion is that the clustering around a seller is negatively assorted. In other words,
    highly connected firms tend to be positioned away from one another
    in the network and thus render the network more vulnerable to
    systemic shock (removal of the highly-connected sellers creates
    an impact across the whole network because their neighborhoods
    are not densely interconnected). The merit here is that the network is less
    likely to percolate in the sense that distress does not spontaneously amplify itself \cite{Newman:2002fk}.

\begin{figure}[h]
   \centering
   \includegraphics[scale=0.35,bb=40 65 450 420]{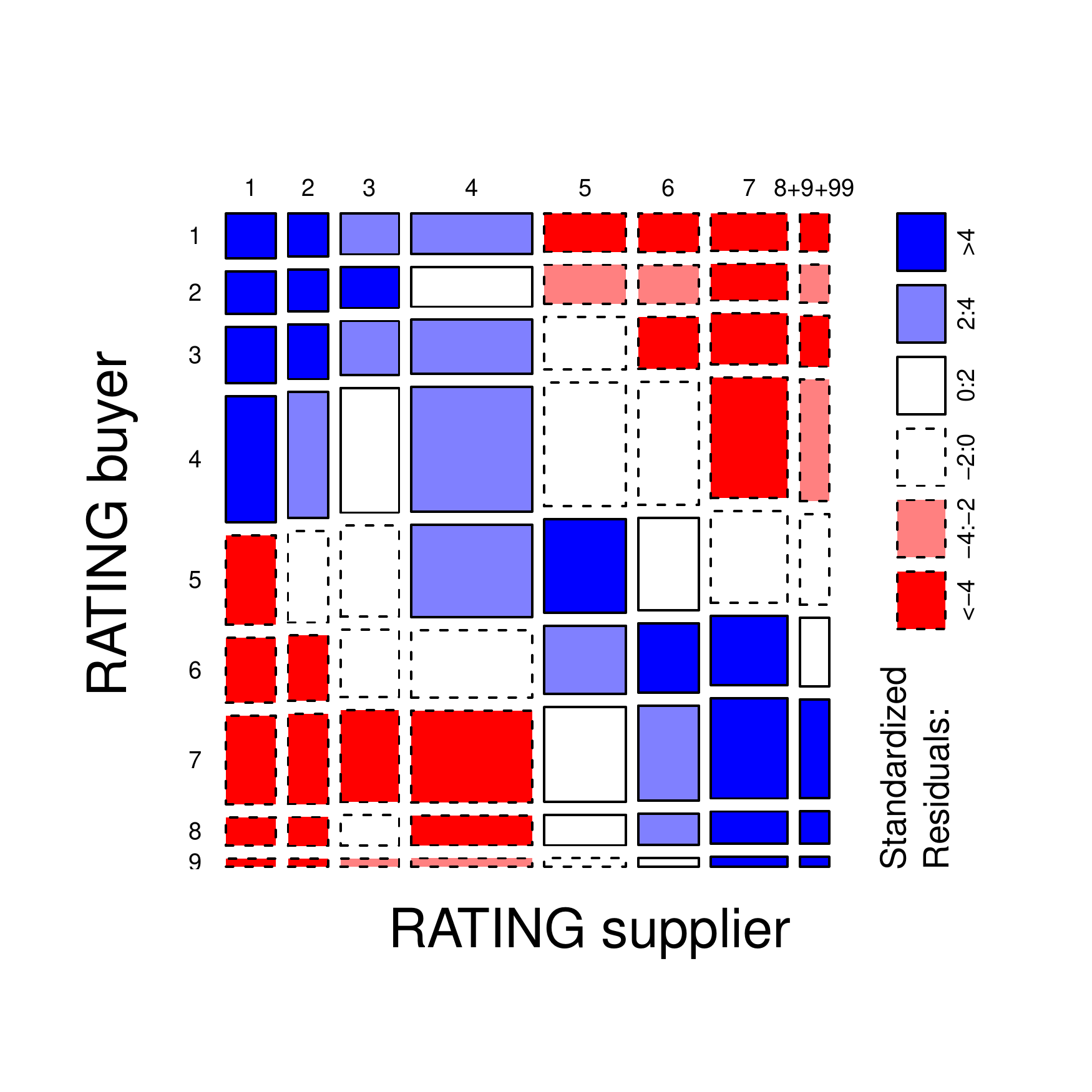}
   \caption{\small Affinity between sellers and buyers.
   A compressed
   tiles mosaic of the cross tabulation RATING seller $\times$ RATING
   buyer of the 2,802,976 $(c,s)$ pairs. The plotting scheme is described in \cite{Friendly:1994uq}. \SRC
   }
   \label{fig:homophily plot}
\end{figure}

 Further, we created a cross tabulation of RATING scores for all
 sellers in the data set $\MAll$ (cf. appendix \ref{app:stats}) and their buyers that have RATING
 information, which is summarized visually in figure~\ref{fig:homophily plot}.
 RATING of the seller is in the columns and RATING of their buyer
 is in the rows. Table elements are therefore enumerations of all pairs of RATING
 scores possible in the data. Essentially this is a description of RATING
 on the two ends of each link between trading firms.

 From this table we created a $\chi^2$ test of independence of
 the categories. This test produced a statistic $\chi^2 =
 2803$ with 56 degrees of freedom and a p-value  identical to zero.
 The conclusion is that we can reject complete independence
 between RATING of a seller and the average RATING of his buyers and suggest a tendency of sellers
 to affiliate with buyers having similar RATING.

  The figure 
   shows the tile mosaic of the paired RATING classes
   with color coding that reveals the tendency
   of sellers to associate with buyers having  a similar RATING;
   the area of each tile in the mosaic is proportional to the
   number of pairs where seller has RATING=X and buyer has RATING=Y.
   A blue tile marks significantly higher than expected
   occurrence, and a red color paints a significantly lower than
   expected pair count.

  The result of combining the stylized facts, and the additional financial parameters leaves a rather empty picture of the full network structure.  Furthermore, the financial insight (like the contract sizes or the production chain lengths) contributes to the uncertainty in classifying the static and dynamic properties of this network. Thus,
   the network measures encourage us to move further to the next step of providing a theory.

   \subsection*{Theoretical foundations for strategic information exposure by different classes of agents}
   \subsubsection*{The principle of thinking at the margin, a profit maximization approach}

    Is there any systematic difference between the amount of information that firms expose to the bank?
  Presenting invoices to stand as collateral against a
  loan with higher face value than is really needed may have a non-trivial effect on the borrower's terms on
  loan: On the one hand, with the now larger collateral presented,
  the face value of the loan will be greater. On the other hand,
  by exposing an invoice representing a sale made to a non credit-worthy
  buyer, the borrower runs the risk of tarnishing their own reputation.%
  \cite{Walther:2008vn}. To better understand the
  workings of these two conflicting effects, we must resort to the
  fundamental economic principle of ``thinking at the margin'', or \emph{marginalism}.
  The application of this principle here is straightforward: once
  enough collateral is discounted to finance the production stage,
  the incremental contribution of yet another invoice in extending
  more credit from the bank becomes unnecessary for covering costs of production.

\begin{figure}[htb]
   \centering
   \includegraphics[page=1,scale=0.31,bb=40 60 460 480]{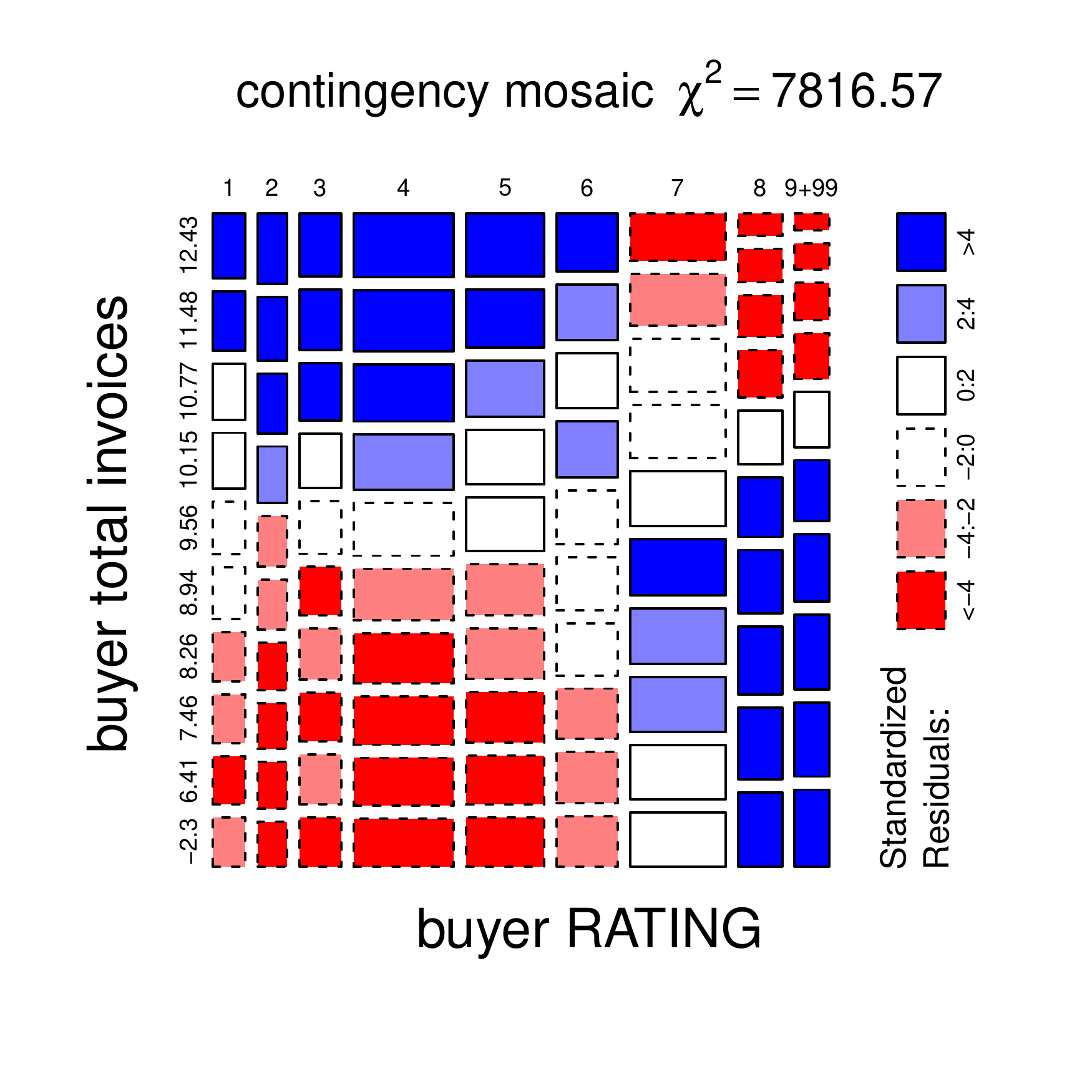}
   \caption{\small A $\chi^2$ test, rejecting complete independence between a buyer firm's credit-rating
   and their total invoiced obligations (p-value $<2.2\times10^{-16}$
   with df=72). The mosaic suggests a tendency of
   the buyers to partition into two groups: credit-worthy buyers
   that pay large amounts in trade-credit, and credit-constrained
   buyers that, in addition, share the low-end of the market purchase power.
   We highlight several points: (1) that middle-rated firms
   exercise purchases to a greater extent than other classes (both high
   and low), and (2) that risky firms make purchases on credit
   at an order of magnitude less than the other classes. }
   \label{fig:BY-BUYER}
\end{figure}

  While this principle of \emph{marginalism} motivates the withholding
  of certain invoices from the bank, there's yet a need to explain
  why this scenario is most appropriate to middle-rated firms. To
  complete the picture, we introduce another well-known concept of
  \emph{Homophily}, which refers to the tendency of agents to
  associate with those bearing similar characteristics. This
  phenomenon was found in many social systems, and many markets in
  particular (e.g. the labor market where profitable firms match
  with high productivity workers, or the marriage market \cite{Becker:1981ve}). As
  seen in Figure \ref{fig:homophily plot}, this is also the case
  in our trade network:
  Sellers associate with same-rating buyers. Now, figure
  \ref{fig:BY-BUYER} provides evidence that low-credit rated buyers
  show a tendency to engage in small sales. This, combined
  with the marginalism argument presented above, has a straightforward
  practical implication: Chances are that credit constrained sellers
  would not sell to large and reputable buyers.
  Should they require financing, they may collateralize the bulk of their invoices, large and small.

  A similar practice is applied by sellers with solid credit histories. These sellers enjoy good terms on loan, so with the aim of getting as much `cheap' credit as possible, these high-rated sellers have no reason to withhold invoices, as the majority of these invoices don't incur the implicit penalty of reduced terms on loan. 
  In between are the middle rated sellers that sell to either high or low rated
  buyers (large and small). These sellers would exercise logic to
  prioritize their receivables when presented for discounting. 

  \subsubsection*{The sensitivity of an agent to its neighborhood}

  Consider a seller embedded inside its trade-neighborhood of buyers
  $\mathfrak{V}_s$. We may recall prerequisite  \textbf{(iii)}
  in the introduction section, that describes
  how the network must lend itself to measure the sensitivities
  of nodes of different species with respect to the neighborhood
  indicator.

\begin{figure}[htb]
   \centering
   \includegraphics[page=1,scale=0.3,bb=15 20 460 450]{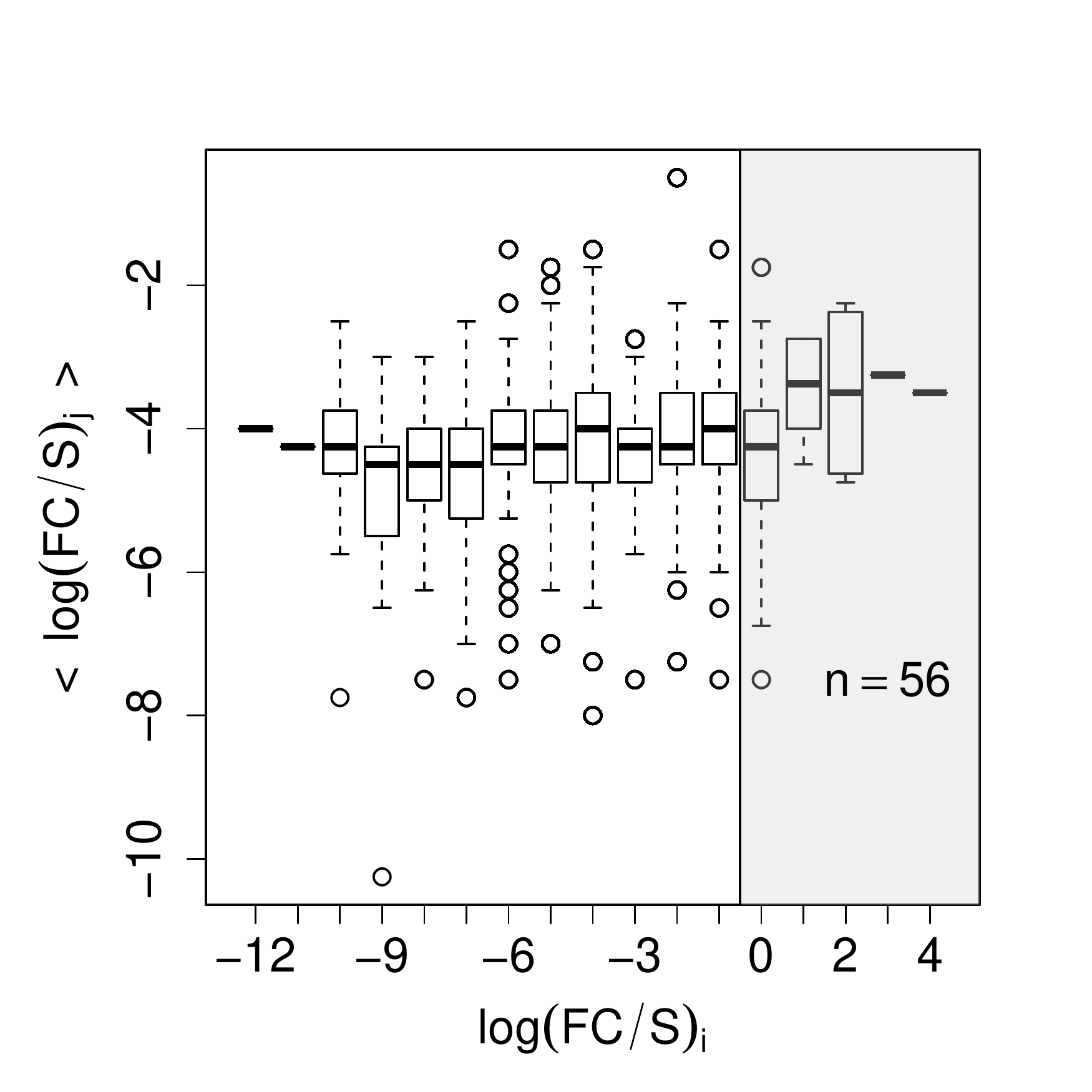}
   \caption{\small Financial costs (interest paid) over sales of a
   seller $\log(FC/S)$ compared against log average FC/S of the
   buyer.  The main effect is negligible between the $FC/S$ of a
   seller and $<FC/S>$ of the trade-neighborhood.}
   \label{fig:FCSperA}
  \end{figure}

  We now consider the financial costs (FC) of the participants.\footnote{Roughly,
  this is the interest paid on loans.}
  This factor is intrinsic to all nodes, including the direct
  neighbors of any agent.
  So it is possible to measure the
  correspondence between an agent's financial costs and
  the aggregated financial costs of its neighbors.
  Figure \ref{fig:FCSperA} plots this estimation. The axes give the
  financial costs of a seller and
  the geometric mean of the financial costs of his customers,
  in logarithmic values. In this instance, little or no
  correlation exists between the financial costs across the trade
  relationships.

\begin{figure}[htb]
   \centering
   \includegraphics[page=2,scale=0.45,bb=15 20 460 450]{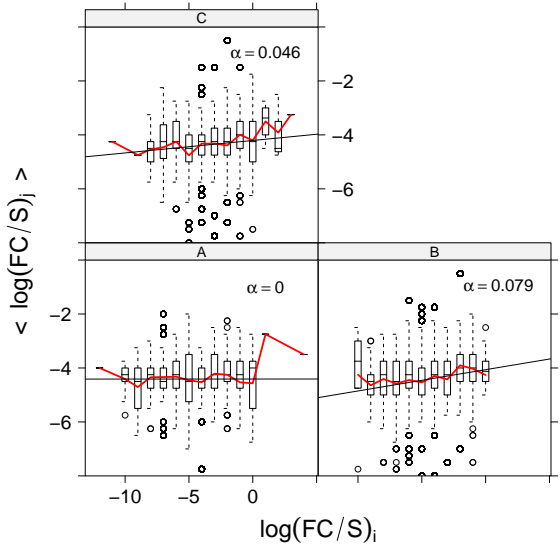}
   \caption{\small with reference to \eqref{eq:neighborhoodIndex},
   $\log(FC/S)$ of the seller is plotted against log average $FC/S$ of the
   buyer neighborhood. The agents are grouped into the three investment grades.
   It is immediately evident that high credit rated sellers (in panel
   A) have little or no correlation with the financial costs of
   their buyers, while the medium (B) and low (C) credit-rated sellers
   percentage increases in financial costs correspond with
   constant percentage changes in the mean financial costs of their
   neighborhood.
   These elasticities are estimated as 7.9\% and 4.6\% respectively.
   }
   \label{fig:FCSperRATING}
  \end{figure}

  We could further plot these
  financial costs vs. neighborhood mean financial costs separately per each credit
  rating class.
  The estimator for different credit rating classes, which is shown
  in figure \ref{fig:FCSperRATING}, gives evidence that a percent
  change in the financial costs of the supplier corresponds with
  up to 7.9\% in the average financial costs across the direct
  neighbors, depending on the credit rating of the seller.
  The greatest effect can be seen among the `B', or medium rated firms, while
  the `A' rated firms show little or no effect. Compared against the overall effect in figure \ref{fig:FCSperA},
   this figure demonstrates that the $FC/S$ ratios of sellers belonging to the `B' and `C' credit rating classes exhibit greater sensitivity to their neighboring nodes.

  It is important at this stage to appreciate that
  different sensitivities to the immediate neighborhood do exist
  through this internal property of financial costs.

  \subsubsection*{The information exposure}

  One of the quantities determining the proportion of the network visible to the bank is
  the ratio
  between the  total amount in the invoices that each seller $s$
  registered at the bank during the year, over their annual net-sales,
  as reported in their financial statement. This provides  the necessary topological aggregate of the network under full information, as required by our method. To describe this
  quantity we use the following parameters.

  We name the agents in the ordered pair $(c,s)$ as the
  (c)ustomer (i.e. the buyer) and the (s)upplier. Both are members
  of the population of agents $V$.

  The amount of invoices that seller $s$ sent to customer $c$ and which he also presented as collateral to the bank for a loan we call $R_{cs}$.
  This is the annual aggregate of the invoices presented by $s$
  on account of the contracts written by him to his customer $c$. The
  face value of an invoice serves as collateral for a short term
  loan in the `credit line'. The annual total of all the invoices that $s$ presented
  to the bank is:
  \begin{align}
    R_s &= \sum_{c \in \mathfrak{V}_s} R_{cs}
  \end{align}
  This amount should be a good proxy for the amount of the total of short term loans
  that $s$ received  to finance production.

   The \textbf{information exposure} of seller $s$, symbolized
  $\am_s$, is the proportion of sales of seller $s$,
  that were presented as collateral:
  \begin{align}
    \am_s = R_s / S_s
    \label{eq:a}
  \end{align}
  A quick interpretation of the information exposure parameter could
  be the following: The net-sales $S_s$ is an item listed in the
  profit and loss statement of a firm and is an annual aggregate
  (also termed a \emph{flow} variable). The net sales item condenses
  all possible profit-making activity of the firm, and implicitly
  includes all the possible information from the firm's trade
  neighborhood in the full network, thus it is a rough description of the reduction in information $\ms \setminus \ms'$.
  The numerator $R_s$, is the aggregate face value of all the invoices presented
  for discount in the short term. Thus, firms running on low profit margins are
  expected to display \a relatively close to 1 because the amount
  of operational credit should be proportional (and close to) the
  sale amount
  . A low ratio of collateral size over net-sales means
  that in the short term, the firm leverages to a lesser extent
  in order to finance the production or goes to get credit elsewhere.
  For this latter claim we assume that our sample is not biased and
  therefore, the credit obtained from other banks occurs uniformly
  across all the firms in our sample.

  The value of \a is greater or equal to zero and can exceed unity.
  There are three possible situations:

  \begin{itemize}
  \item
  \a could be greater than one. A naive\footnote{%
  We assume that our data does not contain traces of illegal
  activity. Otherwise we would have to remove from our dataset firms that have $\am>1$.
  } view for why $\am > 1$ is that there
  is misalignment between the time-frames in the data; the closing of the audit
   and the expiration of all trade-credit contracts
  that were signed in the same year.

  \item
    If seller $s$ has $\am = 0$, the numerator in
    (\ref{eq:a}) vanishes. The interpretation is that $s$ is not a
    direct client of the bank. Rather his customer, $c$, is. The customer
    $c$ entered $s$ into the system by executing an outgoing payment
    transaction.

  \item
  When $\am \ll 1$, a discrepancy exists between the total collateral
  and the net-sales. This could hint that the production of $s$
  requires loans that are smaller than the sales. It is an indication
  of a healthy use of the credit line.
  \end{itemize}

  This variable should not be confused with the information sets
  $\ms$ and $\ms'$. It is however possible that $\am_s$ and $\ms$ may represent the same identities exactly, i.e. in
  networks where only exposure summaries are available. In the example set up, the typical message will be written as
  in \eqref{eq:comprehensiveMS}.
%
  Here $\hat\theta_c$ will be the credit rating of $s$'s customer, $c$.
  Based on the last two sections in the introduction, the RATING score is expected to have
  a non-trivial relationship with information exposure. We will therefore define
  quantities that relate the two, such as the average information
  exposure for each RATING score.
  This is estimated over seller firms $s$ that have rating score
  $r$. The set of sellers with rating $r$ is
  \begin{align}
    \mathfrak{S}_r := \{s:\Rs=r\},
    \label{eq:Sr}
  \end{align}
  where $\Rs$ is the RATING score of firm $s$.
  The average information exposure over all firms having the same RATING is
  \begin{align}
    \bar{\am}(r) = \frac{1}{|\mathfrak{S}_r|}\sum_{s \in \mathfrak{S}_r} \am_s,
    \label{eq:aveaR}
  \end{align}
  where  $\am_s$ is the information set of seller firm $s$.
  In further analysis we will filter out firms having $\am>2$.

   \subsection*{The revealed effect}
\begin{figure}[htb]
   \centering
   \includegraphics[scale=0.32, viewport=5 10 470 450]{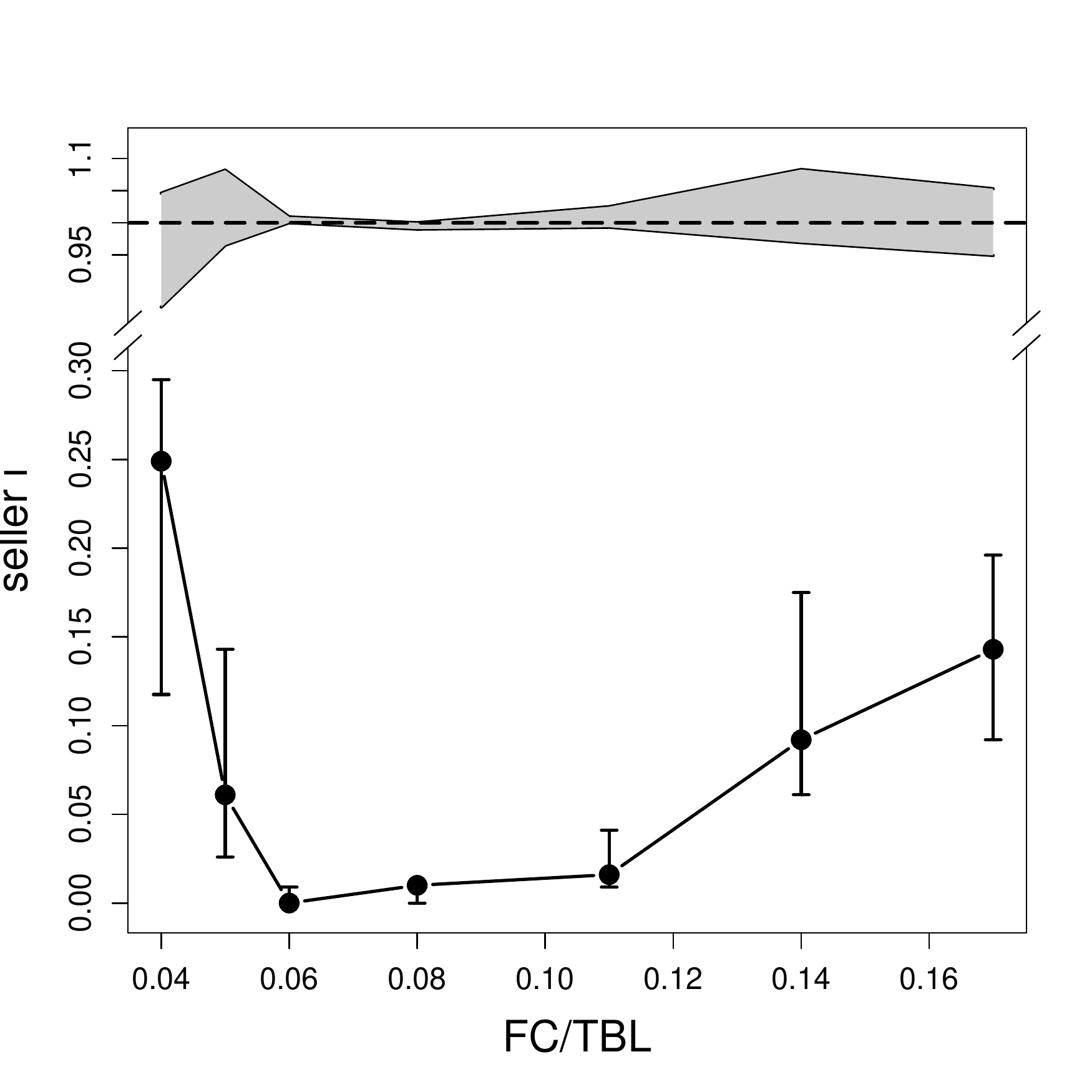}
   \caption{\small Information exposure of sellers against their credit worthiness in two situations: The revealed effect and the expected theoretical effect under the random. To increase our confidence in the ordering of RATING we use Financial Costs over Total Bank Loans as a convenient continuous proxy parameter, since RATING is ordinal (For grouping by RATING classes we refer to fig. \ref{fig:meanAvsRATING}). Far up the Y-scale we theorize the information disclosure under the random, with a symmetric envelope of error bars that have the maximal widths of the main plot on the bottom. There, $\bar{i}$ is approximately unity and independent of the credit-worthiness, so shows \cite{Watanabe:2012fk}.}
   \label{fig:UsVSTakayasu}
  \end{figure}

  Let us consider what would the information exposure parameter $\bar{i}$ look like were we to introduce it under random sampling assumptions.
  For this a work by Watanabe et. al. \cite{Watanabe:2012fk} that estimates the relationship between in-degree and the total of accounts receivable is useful. They used a full business network where only the identities of firms are known (no fund transfers were recorded).
They estimated that under full information $\ms' = \ms$ independent of the rating, and so, grouping \eqref{eq:a} by size classes we have $ \bar{i} = \langle R/S \rangle = \mbox{const} \approx 1 $.
  Figure \ref{fig:UsVSTakayasu} reflects our finding in contrast to this estimation.

  Notably, in the figure we use a continuous quantity as a proxy for credit worthiness otherwise described in this document by the RATING variable. The underlying intuition is straightforward; note that sellers with good credit rating will usually enjoy low interest rates and a relatively seizable credit line. Consequently they obtain low values of FC/TBL. On the other extreme, low-credit firms have limited access to bank credit hence their FC/TBL is expected to be greater ($\chi^2=5522.97$, df=49, p-value $<2.2\times 10^{-16}$, in the test of independece between RATING and FC/TBL).

   As we can see in figure \ref{fig:UsVSTakayasu} the effect of information exposure with credit worthiness is significantly distinguishable from what would be expected under random sampling. This rules out the possibility that random missingness and sampling errors constitute the main effect.

   This is a summarizing evidence in support of the existence of an observer and the implied intentional exposure by the agents. It still remains to see whether other confounding factors may cause this observational equivalence.

   \subsection*{Mediation analysis}
   With reference to figure \ref{fig:homophily plot},
   it is important to note at this stage that
   a $\chi^2$ test of independence is categorical and does not take
   into consideration any ordering of the columns or the rows. However, the
   table used as input maintains the original ordering of
   the RATING classes.
   Thus, the pattern that appears as blue along the diagonal
   does indicate higher-than-expected encounter of similarities in
   the two nodes sharing a trade-link. And any other ordering of the
   columns or rows would result in a less compelling pattern.

\begin{figure}[htb]
   \centering
   \includegraphics[scale=0.3,bb=40 60 480 490]{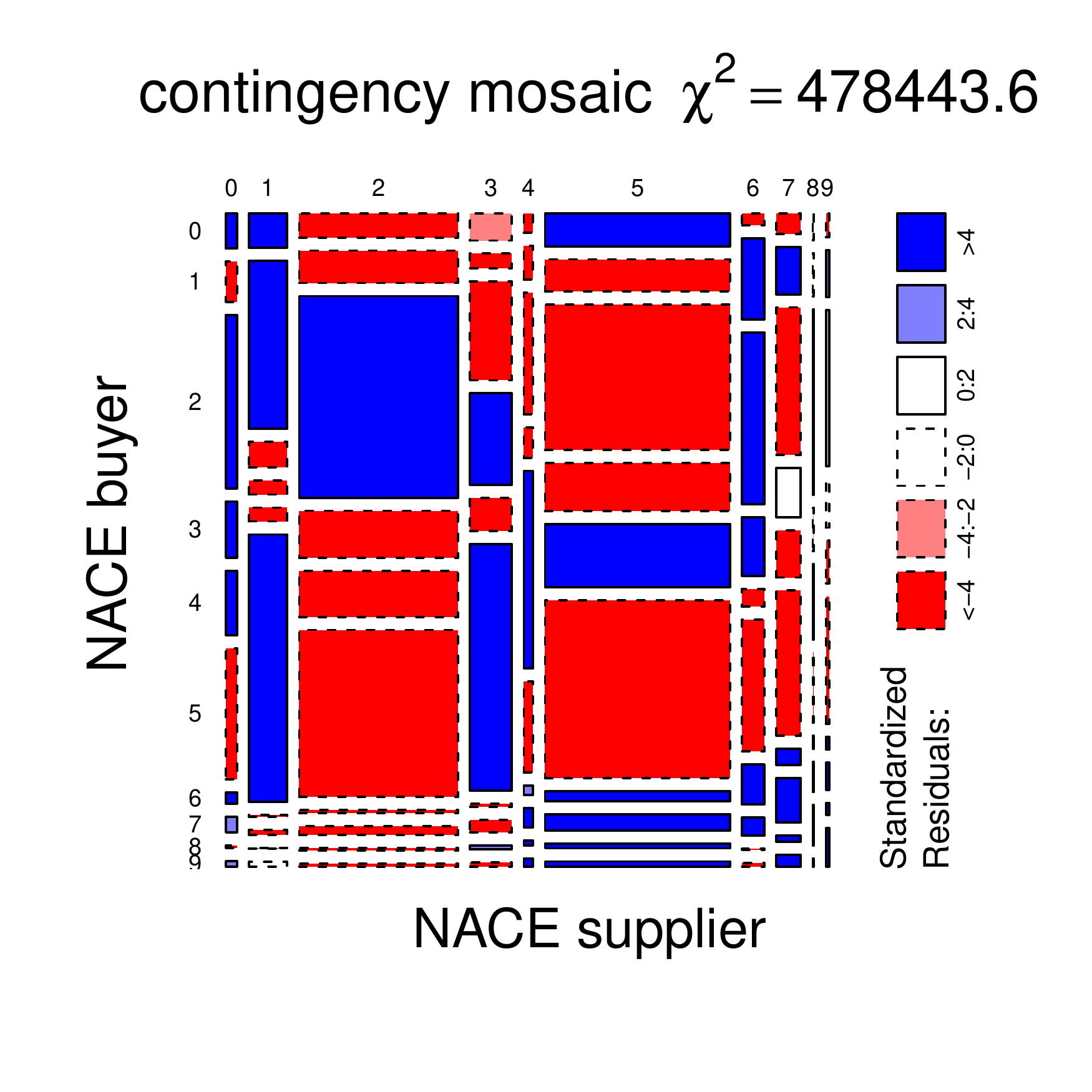}
   \caption{\small Input-Output test of independence between buyer
   and seller. The 1-digit NACE industrial classifications are roughly:
   $1\dots3$ manufacturing, 4 is energy, 5 is wholesale and retail,
   6 is transport, and 7 is real estate. \SRC}
   \label{fig:IO}
\end{figure}

   One reservation could be made on the result above: it is remarkable
   that the RATING of the seller is so similar to the average RATING
   score of their buyers. Looking at the sectoral affiliation of the
   buyers and the sellers (figure \ref{fig:IO}) it seems that
   sellers and buyers are, in the main, trading inside the
   same industries especially when within the manufacturing sectors
   (categories 1 .. 3 of the NACE industrial classification \cite{eurostat:fk}).

   The visual mosaic is symmetric, i.e. it can be
   transposed while maintaining an almost identical pattern of red and
   blue tiles. One exception is major category 5, wholesale, that disrupts this symmetry.
   Firms in industries 1 and 3 sell to firms in 5. According to our
   data set firms in 5 do
   not sell to those in 1 nor 3, but rather to those in 4 (energy).
   In the real world, wholesale trade does connect between manufacturers, but in our case, it only provides a transient path out
   of the manufacturing industries, and splits the supply chain.
   The accounting procedures of wholesale firms are at the root of this problem. A good explanation for this can be found in \cite{Kelman:2015fk}.

   Since sectoral affiliation is known as one of the macro-determiners
   of RATING scores, we addressed this conflation by creating a
   list of firms that trade outside their industry, and then performed
   the test of independence of the RATING score again.
   Approximately 15\% of the buyer-seller links connect firms from
   the same industrial classification.
   Using only the inter-industrial trades, we obtain a similar pattern as in figure \ref{fig:homophily plot} with high degree of confidence.
   Again we can reject the hypothesis of complete independence of the RATING score between buyers and their sellers, and suggest the same trending
   behavior as before.

\begin{figure}[htb]
   \centering
\begin{tikzpicture}[
        every node/.style={anchor=south west,inner sep=-30pt,scale=0.5},
        x=1mm, y=1mm,
      ]
     \node (fig1) at (0,0) %
     {\includegraphics[scale=0.75,bb=10 10 450 440,clip]{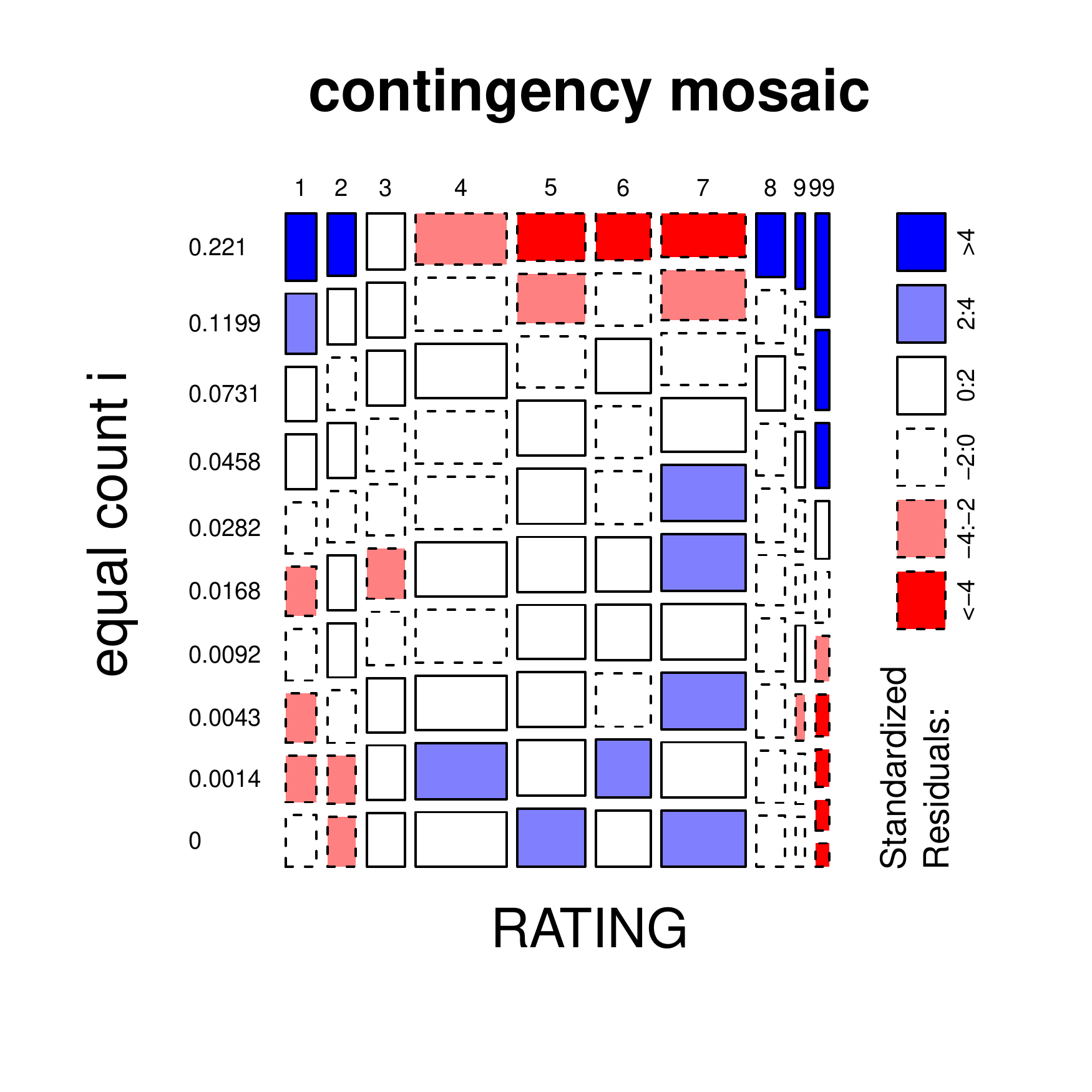}};

     \node (fig2) at (17.5,14)
     {\includegraphics[scale=0.75,bb=90 115 450 300,clip,width=170pt,height=220pt]{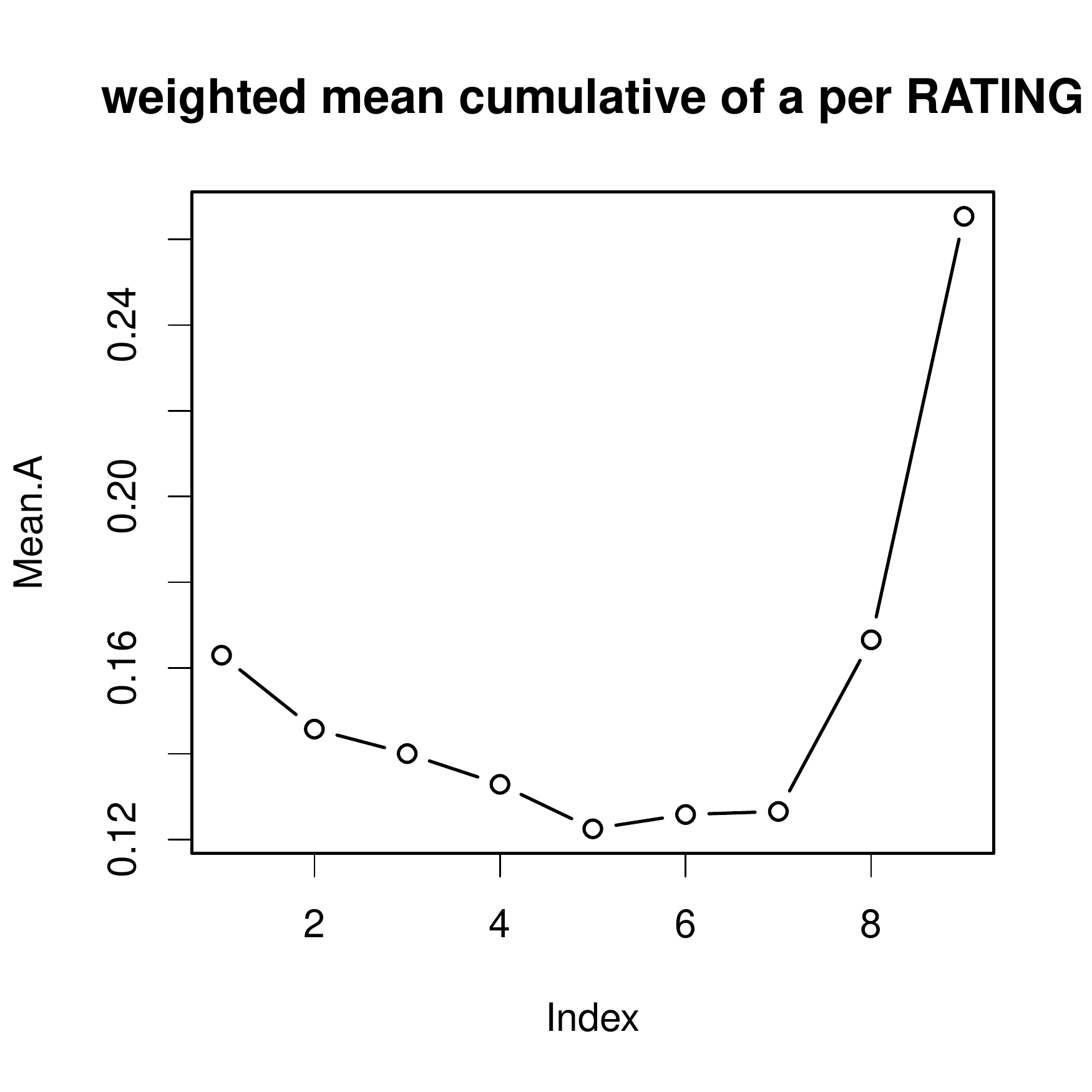}};
\end{tikzpicture}
   \label{subfig:meanAvsRATING mosaic}
   \caption{\small RATING group $r$ vs. average information exposure
   parameter $\bar{\am}(r)$ as defined in (\ref{eq:aveaR}).
   The grouping procedure is described
   in the text. To illustrate the U-shape there is an overlaying line
   plot of the averages $\bar{\am}(r)$. 
   \SRC}
   \label{fig:meanAvsRATING}
\end{figure}

   Figure \ref{fig:meanAvsRATING} gives the relation between RATING
   and the information exposure, \a in the colorful tiles, or its average $\bar{\am}(r)$ as a curve.
   From this figure we learn that
information exposure
does indeed depend
on RATING;  the average information exposure
   is at its minimum in the middle of the RATING scale.  RATING
   scores of firms in the `speculative' financing group $4\dots7$
   have the lowest average information exposure.
  We postulate that these middle-rated firms optimize the amount of information
   they expose.

   To quantify the relation we created a partition
   of the data set into equal-count \ab-groups: the set of 129,584
   suppliers was ordered by \a and then a division into groups was made
   every 12,958 or 12,959 records. The smallest value of \a in each
   group give the tick labels on the Y-axis. 
   The area of each tile in the mosaic is proportional to the count
   of sellers that have RATING=X and $\am \in [Y_t,Y_{t+1})$, $t$ being the tick marker index. The color code marks
   either significantly higher (blue)
   or significantly lower than expected (red) frequency of occurrence.

   The relationship of RATING and \a is, again, a U-shape. 

   In the two extreme RATING scores, 1 and 9, the information exposure is
   the greatest. This sits well with the expectations that `investment' grade
   firms will have dispersed their risk and therefore are
   indifferent to collateral quality. And that firms in risk of default
   will be (or think they are) forced by the bank to surrender all
   possible information.

   It might be that middle-rated borrowers are in less need of credit or get their credit elsewhere.
   However, seeing from the distribution of RATING classes
   most of the firms in the industrial population are
   populating the middle-rated places, and possibly sell to both
   the investment and the credit constrained firms. Given this
   frame, middle-rated firms collect sales (and collateral) from the
   complete range of buyers. Should the ``bad'' buyers coincidentally
   be the ultimate providers of collateral with the smallest total
   face values, prioritizing the invoices is straightforward based
   on probability of payback, and marginal contribution to credit
   financing.

   To show this we present a
   test of independence between buyer's rating class and their total
   purchases on trade-credit (amount of invoices that they received
   from their sellers) in figure \ref{fig:BY-BUYER}.
   The reason not to normalize by the net-sales is that we'd like
   to keep the presented invoices clear of contamination by the
   implicit social factors, and leave in only the direct and visible
   ones.

   We generate a cross tabulation of buyer firm rating class vs.
   equal-counts of buyer firms that are ordered by log of total
   invoice face values. The area (or rather the heights) of the
   tiles in each column are scaled proportional to the relative
   population sizes within the same credit-rating class.

   It is important to note at this stage that from the financial
   statements the proportion of purchases over sales is fixed at
   approximately half (across all financial statements - mean=0.49, median=0.45, n=513,082, sd=0.42
   ).
   In figure \ref{fig:BY-BUYER}, the bulk of firms that make large
   purchases is prominently situated in the middle credit-rated
   classes. As such, it is a tracer for proportionally larger
   production scale of the middle-rated firms and thus they must
   be in greater need for credit, definitely not less.

   Not coincidentally, as the figure shows, if a seller would rank
   buyers according to invoice face values, the incrementally
   small additions to the total sales would be from the lower-end
   rated partners.

   If middle-rated sellers send out invoices exclusively to middle-rated buyers (fig
   \ref{fig:homophily plot}), but middle-rated sellers also need less credit,
   (fig \ref{fig:meanAvsRATING}),
   then middle-rated buyers should also \ul{be purchasing} less in invoices, contrast to what can be seen on figure \ref{fig:BY-BUYER}, where they hold the highest volumes in purchases.

    We could further suggest an alternative scenario that supports the same  observation of the U-shape (figure \ref{fig:meanAvsRATING}), whereby middle credit-rated suppliers use multiple credit sources and  therefore present smaller volumes in collateral to each bank. This may  occur whenever presenting more collateral would have little effect on the terms on  loan with any single bank. Under such `decreasing returns' scenario the U-shape may flatten. However, reaching this plateau warrants starting a new credit line elsewhere. It is noteworthy that this  possibility still falls within our methodology because whether driven by impression management or by decreasing returns, information is withheld due to strategic  considerations on part of the borrower, giving rise to differential information  exposure across species. A risk planning strategy of the bank is considered by us  inferior to the option of active withholding of information by the borrower because competing banks may exercise this same strategy by nature of their  competition, and the fact that borrowers still take their business with the monitoring bank, whose own view this study exposes, means that prioritizing of invoices still occurs.  Controlling for externalities such as competition and geographical proximity to the financial sector is in the scope of another paper currently in preparation.

\section*{Conclusion}
\label{sec:conclusion}

   A common working assumption in many studies 
   is that the observed network is observed at random and that the network available to us is a representative
   sample of the true network. A major challenge to this view is
   posed by the possibility that the network is monitored.
   Not only is the true network not visible to us, neither is the
   process that causes the data to be missing.
   So a direct statistical test is
   not feasible. We offer to circumvent this problem by a cascade
   of hypotheses and mediation analysis.
   Assuming the network is not monitored corresponds with the
   fact that the observer's payoff is not sensitive to the quality
   of the social ties, and the agents are not expected
   to withhold information on quality. Therefore, the information
   exposed by agents in the unmonitored network should be
   independent of the species of the agent or else a spurious correlation should exist owing
   to the intervention of mediators. Controlling for these mediators,
   one can, in theory, test whether the network available to us
   possesses the species-independence property or not.



  In the example we were able to: \textbf{(a)} show that data are missing but not
  at random and suggest that the data collector has an impact
  on the process by which firms appear to associate, and especially
  on a certain subgroup of the agents, namely the medium
  credit-rated firms;
  \textbf{(b)} focusing on this subset of the agents, there is a clear social dimension which is related to an accounting behaviour
   that is fondly termed in the literature `window-dressing'
   \cite{Allen:1992fk} and thus
  \textbf{(c)} show that the immediate neighborhoods of these agents
  are more likely to be missing complete links and nodes structure;
  \textbf{(d)} suggest a tendency of this network to be vulnerable
  to targeted removal of specific species of nodes. These species may be
  correlated with industries, similar to the scenario of a systemic
  shock arising when the bank issues a regulatory action on the industry
  as a whole.


  Since the bank is collecting the data and so assumes a point of view
  we suggest that \textbf{(e)} 
  distress response due to internal dynamics is rarely visible to
  the bank. The synchronized response subsequent
  to an intervention of the regulatory system is 
  visible, however, because of its causal nature.

  \section*{Implications, applications and limitations}
  \label{sec:implications}


  We have developed a method for exploring  data on networks in situations where hidden information affects the outcome.
   Essentially to measure a network under the hypothesis that data are missing not at random, we choose a set that has \textbf{(a)} a system of  interacting agents classifiable into species, \textbf{(b)} an observer and visible interactions between it and the players, \textbf{(c)} a theory that describes the payoff and offers interpretation to why would different species play different strategies. Finally \textbf{(d)} a measurable quantity of information that can be compared from additional input.

  The proposed methodology can be used to derive the likelihood
  that an agent embedded in a network hides information conditional
  on some characteristic of the agent. In our empirical example this
  likelihood is associated with the credit worthiness of an industrial firm: medium rated
  firms have a higher probability to strategically hide information
  from the observer. The observer, in our specific case a
  bank, can use the proposed methodology to identify a subset of
  the population of corporate customers that are highly likely to
  hide information. This subset can be the target of a specific
  action by the bank aimed at probing the supply chain connections
  of the firm that are currently hidden. For example, a possible
  policy would be to offer a lower interest rate on loans if the
  firms present an invoice as collateral. In a simple version this
  reduction can be flat regardless of the rating of the buyer that
  is listed on the invoice. In a more aggressive version the
  bank can offer a higher reduction on interest for presenting invoices
  issued to lower rated buyers. The goal is to create an incentive
  scheme that encourages the suppliers to disclose invoices
  that they otherwise would be inclined to hide. Once the bank has
  acquired a more precise picture of the firms' supply chains, it
  can adjust the rating of the firms accordingly. The proposed
  algorithm is essential in limiting the subset of firms that are
  the target of these costly actions.

  We believe that there is an internal ranking mechanism of social tie quality employed
  by agents that renders their incentive to expose information about
  some peers but not others. 
  In trade networks 
  the lower ranked nodes may
  threaten the stability of the structure more than the better-ranked
  ones. Therefore, the more links are missing, the more is the structure
  susceptible to dynamic breakage. This intuition encourages the
  possible replication of the method in other fields where networks
  have a social component with either a ranking mechanism, or
  that information exposure could be quantified.


  Last, this study points out the possibility that a single observer
  exists and that this entity not only monitors the agents but affects their payoffs.
  In
  other networks, several observers may exist and so agents may
  vary their strategies accordingly; thus further concealing their links
  or even creating links that do not exist. 

\section*{Acknowledgement}
  We acknowledge the partial support by the Institute for New Economic Thinking (INET), grant IN01100017.

\bibliographystyle{ws-acs} 
\bibliography{biblio_main}

\onecolumn

  \begin{appendix}
\section{Financial trade networks}
\label{app:financialNetowrks}
   Financial networks are known for being negatively assorted,
   i.e. neighbouring nodes in the network are dissimilar, in
   particular as regards to the  degree of their in- and out-links.
   Among practitioners and economists this property is desired because
   it renders the financial network robust to percolation (propagation
   of distress or growth). The knowledge that contagion rarely
   happens may catch us by surprise when financial shocks
   do indeed propagate from the local level to the national/international
   level. In the events preceding the 2008 financial crisis, small
   systemic shocks affected large proportions of the industrial and
   trade networks. The usual response of firms to market downturns
   was then amplified and this response swept across the network using
   the monetary (communication) channels. One reason for the lack of
   control over this incident was that a proportion of the communication
   channels was not known to the banking
   system: the high risk mortgages were traded in the market but
   the credit-unworthy clients behind them remained anonymous.

   \subsection{General characteristics of trade networks}
    Assortative mixing in networks is a term describing the correlation
    of `popularity' between different nodes. Popularity is attributed
    to a node and measured by the number of incoming links to it.
    A network is positively assorted if the number of incoming links
    to a node is positively correlated with the number of incoming
    links of its neighbors. In assorted networks messages can spread
    within a small number of steps since there are many redundant
    links via which a message could travel.  Negatively assorted
    networks contain highly connected nodes that are positioned
    sparsely throughout the network. Thus, in this species of topology
    the fast spread of messages is less likely \cite{Newman:2002fk}.
    However, if attributes of nodes are known, it is possible to
    combine structural and behavioral information for efficient
    routing inside this network \cite{Simsek:2008kx}.

    \subsection{Characteristics of the network under study}
   The asymmetric links in this network
   represent the exchange of goods/services for financial payments
   between agents; it is similar to the better known
   communication networks.
   Each link is between a seller firm and one of its customers who bought a product or service from them.
   Our data contain a snapshot in time of many firms that provide
   goods and services in exchange for financial payments in the
   year 2007. This was the year when financial crises were occurring
   global-wide \cite{Duchin:2010bh}.

   These data were collected by a single large Italian bank,
   fulling its function as an intermediating agent in a delayed
   payment procedure. The bank recorded the names of the two parties
   and the amounts of money that one firm, the buyer, owes the
   other, the seller.

  In the data, the network contains a record of financial interactions between peers.
  The interaction under investigation is recorded when a \emph{discount}
  process occurs. The bare explanation of a discount on an invoice
  is that an owner of an invoice will sell it to a financial institute
  for a lower price than its face value. The buyer of the invoice
  will be the new creditor and will take upon himself the risk that
  the debtor will become insolvent.  This risk is combined into the
  rate of discount.

  Today, banks offer their customers a cheaper alternative to selling their
  trade bills. A customer of the bank can `collateralize its accounts
  receivable': instead of buying the invoice, the bank will extend
  a secured loan using the face value of the invoice as collateral.
  It is commonly known that, when using discounted invoices, the
  borrower firm is a
  seller, not a buyer. Extensive reviews of the reasons why
  this may be so were suggested in \cite{Omiccioli:zr}
  and \cite{Marotta:2000ys}.
  Here we note one obvious reason: the seller needs to
  secure funds only for production of the goods/services, whereas the buyer
  needs to cover the total amount of the invoice. i.e.
  the costs of production \ul{and} the seller's profit.  Lower amounts
  on loan impose less risk on the lender and in return a more
  affordable discount is offered.

  The data are not publicly available; they were directly accessible
  to only one researcher, who worked for the bank.  Programs to
  extract summary data from these two data sources were written
  by us and executed on a computer inside the bank.
  The summary data, which we possess, were then further analysed
  to obtain the results reported here.

%

  \section{Credit rating and financial costs}
  \label{app:creditRatingCosts}
  In order to facilitate an efficient discounting mechanism the
  banks created a credit-rating procedure. When the customers of the
  bank require loans, they should qualify as credit-worthy, i.e. be
  borrowers that are financially capable of paying back. Credit-rating
  is a score provided to all banks by an external entity;\footnote{CeBi
  - Centrale dei Bilanchi, a financial analyses service for the
  Italian banking system.}
  that uses a computerized system
  which automatically assigns a RATING score, $\Rs$, to each firm.

\begin{figure}[htb]
   \centering
   \subfigure[frequency]{
   \includegraphics[clip, scale=0.32,bb=5 70 438 415]{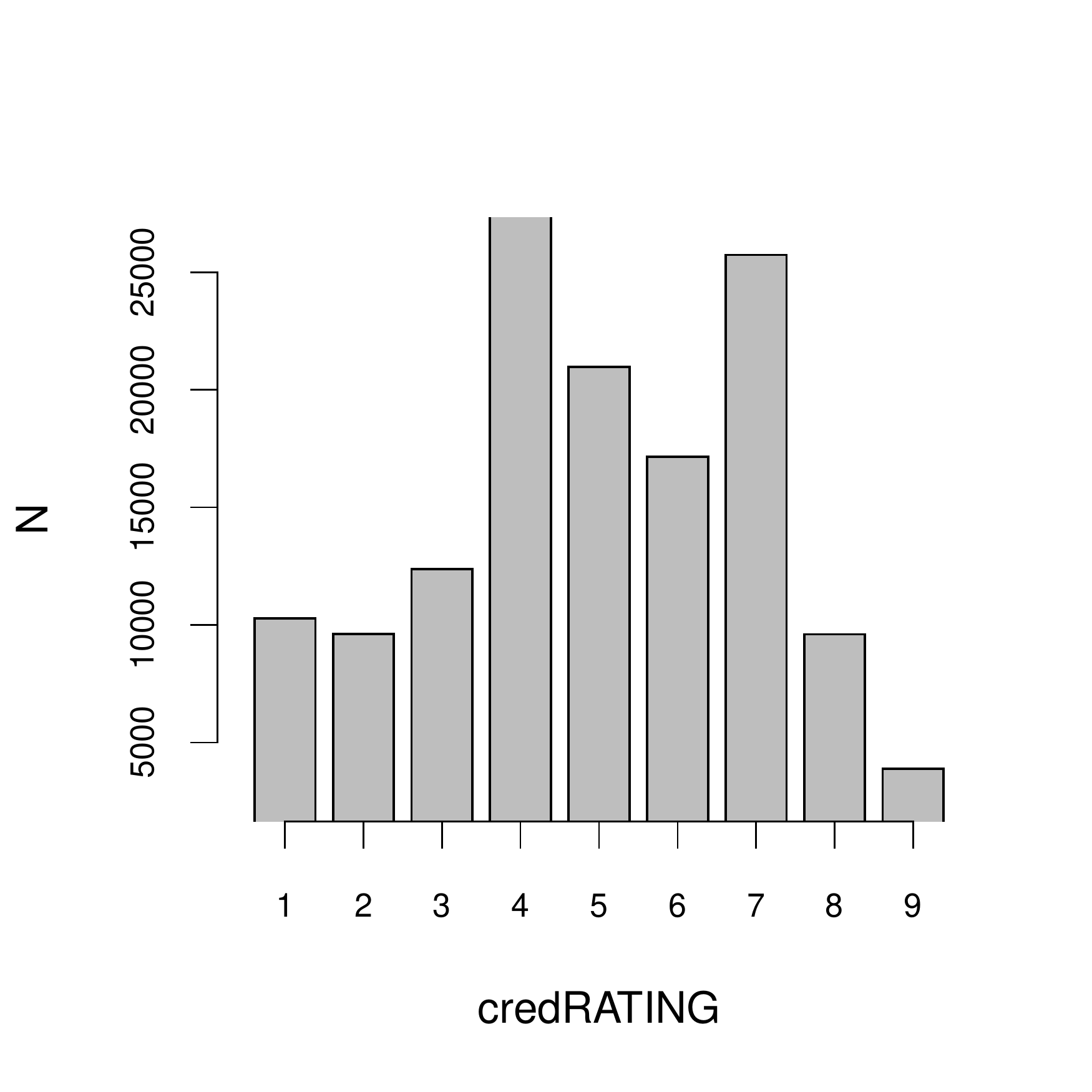}
   \label{subfig:histNumRATING}
   }
   \subfigure[size]{
   \includegraphics[clip, scale=0.32, bb=5 70 457 415]{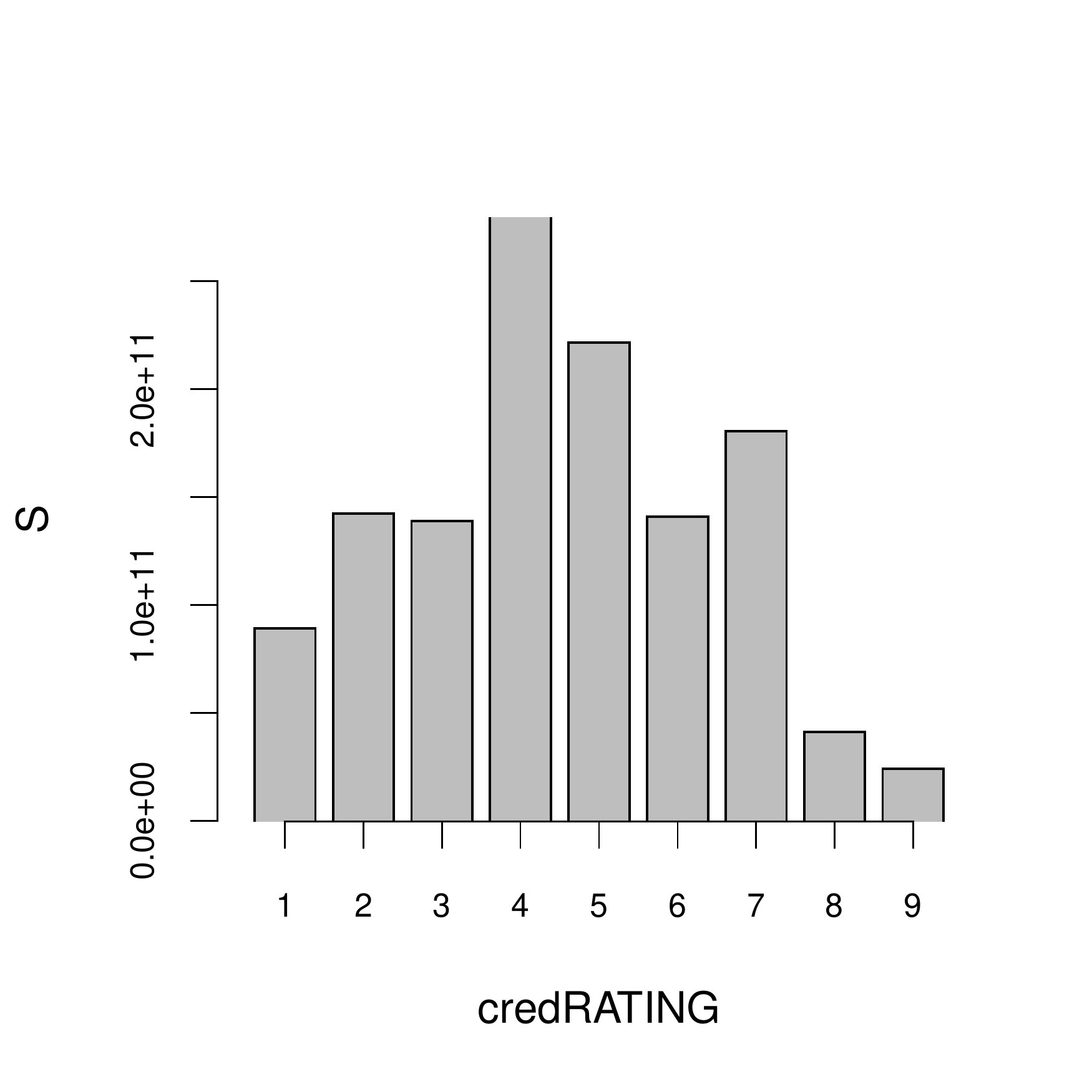}
   \label{subfig:histSizeRATING}
   }
   \caption{\small Histograms of sellers in $\MAll$ (cf appendix \ref{app:stats}): firm counts
   (N) and total net-sales (S) per RATING score.
   RATING=1, 2, 7 and 8 exhibit the largest deviation of sales per number of firms. \SRC
   }
   \label{fig:histRATING}
\end{figure}

  The score is independently computed from the financial statements
  of each firm and is in the range $1\dots 9$ where low credit-rating
  is
  indicated by a high value. The common practice is to further group this index into
  classes: high ($\Rs = \{1\dots3\}$) which characterizes firms
  of investment grade, medium  $\{4\dots6\}$ for firms
  somewhat susceptible to defaults, and low credit rating $\{7\dots9\}$.
  Firms
  that score into the `low' class have a high probability to default
  on payments. They are regarded by practitioners as having little
  or no access to bank credit.  A firm with a score of 9 will rarely
  qualify for borrowing. However, since these firms appear in our
  data as borrowers from the bank, we assume that they did receive
  loans (figure \ref{fig:histRATING}); low credit rating could be
  caused by the industry in which the firm operates.
  The calculation of the RATING score is proprietary but shows correlation with
  Altman's Z-score \cite{Altman:1968ys}. For a comprehensive explanation
  of the RATING score the reader is referred to \cite{Bottazzi:2011fk}.
  It is also important to note that the RATING score of a firm
  is visible to all the banks with whom that firm does business.
  This is part of a transparent national credit system that was
  erected in Italy. A common credit registry is also available in
  other countries.


  Naturally, there are more buyers than sellers. However, the
  distribution of RATING scores is identical once normalized by the
  total number in each group. Figure \ref{fig:histRATING} displays
  the RATING histograms of sellers.

  Credit-rating affects the terms on loans.
  When a firm believes that its bank is imposing conditions that
  are unreasonable, it may resort to other means of financing.
  In general, banks are the sole providers of loan financing and
  by declining a loan based on its terms (interest rate)
  the firm must consider other financing channels, the most intuitive
  of which are dropping the contracts, or using trade-credit; the firm will ask to delay its
  debt to sellers and collect immediate payments from its buyers
  \cite{Petersen:1997kx}.
  In either case, a social component is added to the pool of financing
  channels, and this can be traced on the network of trading firms.

  \section{Fundamental statistics}
  \label{app:stats}
   The privately operated and public firms in the Italian industries are all borrowers from the bank. The data sets contain individual firm level information about transactions between firms (registered, collateralized invoices) and financial statement data.
  In the example there are 1,578,812 firms connected
   by 7,290,072 trade links but only 273,726 of them are sellers (have
   incoming trade-credit links).
   The network nodes are crossed with
   available balance sheet data of 703,858 firms in 2007. The balance
   sheet data come from the Italian Centrale dei Bilanchi (CeBi)
   and is similar in structure to \texttt{MICRO.1} that was presented
   in \cite{Bottazzi:2006fk} and elsewhere.
   The resulting network has 345,403 nodes connected
   by 2,874,830 links.
   In this network, 140,580 nodes are sellers
   and 129,584 of them are sellers that appear in the next time.
   This subgroup of sellers is named $\MAll$ and the coverage by industry appears in table \ref{tab:coverage} below. It is notable that the manufacturing and construction are the largest industries.
   Being a seller in $\MAll$ means having balance sheet
   sales that are greater than zero, at least one incoming payment link,
   and the linked buyers also have balance sheet information.
   Importantly, 122,728 of the sellers in $\MAll$ (94\%)
   have additional outgoing links, so they are also buyers. The
   remaining 215,819 firms in the network appear only as buyers.
   Thus, to a proportion of one half the network of buyers
   and sellers cannot be classified as bipartite since breaking
   the links from exclusive buyers leaves one third of the network
   intact ($1/2 \cdot 2/3 = 1/3$).
   Further, this observation design is
   at random:
   there are two bank consortia in Italy, and any borrower may do business with one or the other.
   The nodes in $\MAll$ are
   guaranteed to be customers of this bank (borrowers), but a similar number of customers should therefore be of the other bank. The remainder are simply buyers that do not finance production via trade-credit.
   So we should expect an overlap of one third of the firms in the records of both banks. Indeed, all nodes are buyers but out of 345,403 in total,
   62\% are not sellers that borrow from the bank that collects the data.


   \renewcommand{\arraystretch}{1.}
\begin{table*}\par
\begin{minipage}[t]{0.25\linewidth}\centering
	\begin{tabular}{c|c}
NACE&N \\
\hline
10 & 3  \\
11 & 426  \\
12 & 84  \\
13 & 9  \\
14 & 757  \\
15 & 3073  \\
17 & 3779  \\
18 & 2160  \\
19 & 2076  \\
20 & 2102  \\
21 & 1433  \\
22 & 3121  \\
23 & 160  \\
24 & 2438  \\
25 & 3959  \\
26 & 3444  \\
27 & 1367  \\
28 & 12628
      \end{tabular}
  \end{minipage}%
\hfill%
\begin{minipage}[t]{0.25\linewidth}\centering
	\begin{tabular}{c|c}
NACE&N \\
\hline
29 & 9199  \\
30 & 298  \\
31 & 2730  \\
32 & 726  \\
33 & 1721  \\
34 & 771  \\
35 & 414  \\
36 & 4744  \\
37 & 304  \\
40 & 42  \\
41 & 12  \\
45 & 7997  \\
50 & 4818  \\
51 & 27808  \\
52 & 3843  \\
55 & 194  \\
60 & 2550  \\
61 & 21
      \end{tabular}
  \end{minipage}%
\hfill%
\begin{minipage}[t]{0.25\linewidth}\centering
	\begin{tabular}{c|c}
NACE&N \\
\hline
62 & 7  \\
63 & 1360  \\
64 & 135  \\
65 & 1  \\
67 & 51  \\
70 & 87  \\
71 & 382  \\
72 & 3731  \\
73 & 49  \\
74 & 4636  \\
80 & 92  \\
85 & 220  \\
90 & 556  \\
91 & 10  \\
92 & 273  \\
93 & 214  \\
 &   \\
 &   \\
      \end{tabular}
  \end{minipage}%
  \caption{Sample coverage: the number of seller firms in the trade network of the year 2007, sorted by their industrial sector. The list above excludes 6,569 firms with unidentified sectors. The industrial codes follow the format in NACE v1.1\cite{eurostat:fk}. Roughly 10-13 are mining, 15-37 are manufacturing, 50-52 is construction, and 71-74 are services.}
  \label{tab:coverage}
\end{table*}
NACE - The European industrial classification scheme is a   hierarchical numbering system. The leftmost digit is the major industry code. Further sub-classifications can be achieved by inspecting the less significant digits, up to 4 digits. In our analyses we use a two-digit classification.
   \end{appendix}

\end{document}